\documentclass[12pt]{article}
\pdfoutput=1
\usepackage{amsmath,amssymb,amsthm,amscd}
\usepackage{tikz}
\usepackage{jheppub}
\usepackage{amsmath}
\usepackage{enumerate}
\usepackage{placeins}
\usepackage{array}

\theoremstyle{definition}

\def\e{\textrm{e}}

\newcommand{\bea}{\begin{eqnarray}}
\newcommand{\eea}{\end{eqnarray}}
\newcommand{\be}{\begin{eqnarray}}
\newcommand{\ee}{\end{eqnarray}}
\newcommand{\nn}{\nonumber}
\newcommand{\Tr}{\mathrm{Tr}}

\newcommand{\udl}[1]{\mathrm{d} #1 \,}

\newcommand{\qfac}[1]{\left( #1; q \right)_\infty}
\newcommand{\xfac}[1]{\left( #1; x^2 \right)_\infty}

\newcommand{\acom}[2]{\left\{ #1, #2\right\}}
\newcommand{\sbfunc}[1]{s_b\left( #1\right)}
\newcommand{\Sfunc}[1]{S_2\left( #1\right)}
\newcommand{\SFunc}[1]{S_3\left( #1\right)}

\def\ga{\alpha}
\def\gb{\beta}
\def\gc{\gamma}
\def\Gc{\Gamma}

\def\gd{\delta}

\def\gt{\theta}

\def\gs{\sigma}

\def\gr{\rho}
\def\gp{\phi}
\def\Gp{\Phi}

\title{From \boldmath$3d$ dualities to $2d$ free field correlators and back}

\abstract{We investigate the relation between $3d$ $\mathcal{N}=2$ theories and $2d$ free field correlators or
Dotsenko-Fateev (DF) integrals for Liouville CFT. We show that the $S^2\times S^1$ partition functions of some
known $3d$ Seiberg-like dualities reduce, in a suitable $2d$ limit, to known  basic duality identities for  DF correlators. These identities
are applied in a variety of contexts in CFT, as for example in the derivation of the DOZZ 3-point function.
Reversing the logic, we can try to guess new $3d$ IR dualities which reduce to more intricate duality relations for the DF correlators.
For example, we show that a recently proposed duality relating the $U(N)$ theory with one flavor and one adjoint to a WZ model can be regarded as the $3d$ ancestor of the evaluation formula for the DF integral representation of the 3-point correlator.
We are also able to interpret the analytic continuation in the number of screening charges, which is performed on the CFT side to reconstruct the DOZZ 3-point function, as the geometric transition relating the $3d$ $U(N)$ theory to the $5d$ $T_2$ theory.}
\author[1]{Sara Pasquetti}
\author[1]{Matteo Sacchi}

\affiliation[1]{Dipartimento di Fisica, Universit\`a di Milano-Bicocca \& INFN, Sezione di Milano-Bicocca, \\
I-20126 Milano, Italy
}

\emailAdd{sara.pasquetti@gmail.com}
\emailAdd{m.sacchi13@campus.unimib.it}

\begin{document}

\maketitle

\section{Introduction}

In recent years the connection between $3d$ $\mathcal{N}=2$ gauge theories and free field correlators or Dotsenko-Fateev (DF) blocks in Toda CFT has been considered from various perspectives. In \cite{Aganagic:2013tta,Aganagic:2014oia} a dictionary to map $3d$ quiver theories to $q$-deformed DF blocks was proposed. This map was further explored on various compact spaces in \cite{Nedelin:2016gwu} and from a different perspective in  \cite{Kimura:2015rgi}.

On the other hand, it has been recently observed that a correspondence between $2d$ quiver theories and DF blocks can be obtained from the reduction of $3d$ dualities of spectral or mirror type  \cite{Zenkevich:2017ylb}.

The reduction of $3d$ dualities to $2d$  is quite delicate, as discussed extensively in  \cite{Aharony:2017adm}. To avoid subtleties with non compact directions we can focus on the case where all the mass deformations are turned on, so that the theories have isolated vacua.
When the theory is put on a space with a compactified dimension, such as $D^2\times S^1$ or $S^2\times S^1$, we have to specify how the parameters, namely real masses and FI parameters, scale with the radius of the $S^1$ as it goes to zero. 

One possibility consists of the \emph{Higgs limit}, where the matter fields remain light while the Coulomb branch is lifted.
 Alternatively, one can consider the opposite limit, called \emph{Coulomb limit}.
 At the level of the partition function, the two limits are implemented by the following properties of the $q$-Pochhammer symbol, in terms of which the contribution of a $3d$ chiral field is written,
\be
&&\lim_{q\rightarrow 1}\frac{\qfac{q^x}}{\qfac{q^y}}=\frac{\Gc(y)}{\Gc(x)}(1-q)^{y-x}\nn\\
&&\lim_{q\rightarrow 1}\frac{\qfac{zq^x}}{\qfac{zq^y}}=(1-z)^{y-x}\, .
\ee
The first identity is used when considering the Higgs limit, while the second is used for the Coulomb limit. Hence, the Higgs limit typically results in the partition function on $D_2$ or $S^2$ of a GLSM, where the contribution of a $2d$ chiral is written in terms of gamma functions. Instead, the result of the Coulomb limit with some manipulations can be mapped to the partition function of a Landau-Ginzburg model with logarithmic twisted superpotential. 

If we have a mirror or spectral dual pair, since  Coulomb and Higgs branches are swapped, if we take the Higgs limit on the electric side we are forced to take the Coulomb limit on the magnetic dual side.
This is how  $3d$ Mirror Symmetry  reduces in $2d$ to the Hori--Vafa duality \cite{Aganagic:2001uw}.

However, as a matter of fact the integrals that one gets in the Coulomb limit are directly mappable to DF integrals. In particular, in \cite{Zenkevich:2017ylb} it was suggested that taking the $q\rightarrow 1$  limit of a spectral dual pair \cite{Aprile:2018oau} yields a relation between a $2d$ gauge theory and DF conformal block. For example, in the case of the $FT[SU(N)]$ theory one gets the integral block for two arbitrary and $N$ degenerate primaries in $A_{N-1}$ Toda theory. This is basically the $2d$ version of the map by \cite{Aganagic:2013tta,Aganagic:2014oia} between a correlator with degenerate operators (represented as a screening integral) and the vortex partition function of $2d$ surface operator.

Another possibility is to consider the $2d$ limit of $3d$ Seiberg-like dualities, such as the Aharony duality \cite{Aharony:1997gp} and some variants with monopole superpotential discussed in \cite{Benini:2017dud}. In these cases, Higgs and Coulomb branch are not exchanged by the duality. Hence, we can take the Higgs limit on both sides of the duality, obtaining a  duality between GLSM's which is a sort of $2d$ version of the Seiberg-like duality we started with. Conversely, we can consider the Coulomb limit of the dual theories, which will result in some identities between DF integrals. 
 
Indeed we show that by carefully taking the Coulomb limit of the superconformal index, that is the partition function on $S^2\times S^1$, for the monopole dualities \cite{Benini:2017dud}, we land on the {\it duality  relations} between DF conformal correlators that can be found for instance in \cite{Fateev:2007qn,Baseilhac:1998eq}. 

One can then try to reverse the process: is it possible to {\it uplift} some known dualities between DF integrals in $2d$ CFT to genuine (IR, not mass deformed) dualities between $3d$ gauge theories? This is the philosophy of this paper and its companion \cite{PS}. More precisely, what we are doing is not reversing the RG flow, but guessing what is the $3d$ gauge theory ancestor of the $2d$ relation between DF correlators.

The first example where we are able to perform such an {\it uplift} is what revolves around the evaluation of the 3-point function in $2d$ Liouville CFT. It was observed in \cite{GL} that by integrating over the Liouville zero modes, correlators of $k$ primary operators develop poles when the momenta satisfy the screening quantization condition
\begin{equation}
\ga\equiv\alpha_1+\cdots +\alpha_k=Q-Nb, \qquad N \in \mathbb{N}\, ,
\label{onshell}
\end{equation}
where $Q=b+b^{-1}$ and $b$ is the coupling constant appearing in the central charge
$c=1+6Q^2$.
The residue in turn takes the form of a free-field Dotsenko-Fateev (DF) correlator with $N$ screening charges.

In the case of the 3-point correlator yielding the structure constant $C(\ga_1,\ga_2,\ga_3)$, we have
\begin{equation}
   \underset{\alpha=Q-Nb}{\text{res}}C(\alpha_1,\alpha_2,\alpha_3)=
   (-\pi\mu)^N
   I_N(\alpha_1,\alpha_2,\alpha_3)\, ,
\end{equation}
with 
\begin{equation}
   I_N(\alpha_1,\alpha_2,\alpha_3)=
   \int\prod_{j=1}^N |t_j|^{-4b\alpha_1}|t_j-1|^{-4b\alpha_2}
      \prod_{i<j}^N|t_i-t_j|^{-4b^2}
   \,\udl{^2\vec{t}_N}
   \label{IN}
\end{equation}
and
\begin{equation}
\udl{^2\vec{t}_n}=\frac{1}{\pi^nn!}\prod_{k=1}^n\udl{^2t_k}\,.
\end{equation}
The integral \eqref{IN} was calculated exactly in \cite{Dotsenko:1984ad} and then used to guess the form of the 3-point function via analytic continuation, as we will review. However, in \cite{Fateev:2007qn} a different derivation of the evaluation formula of \eqref{IN} was provided, based on the duality relation for the DF integrals \cite{Baseilhac:1998eq}:
\be
&&\int\prod_{i<j}^{N_c}|y_i-y_j|^2\prod_{i=1}^{N_c}\prod_{a=1}^{N_f}|y_i-\tau_a|^{2p_a}\udl{^2\vec{y}_{N_c}}=\frac{\prod_{a=1}^{N_f}\gc(1+p_a)}{\gc(1+N_c+\sum_ap_a)}\prod_{a<b}^{N_f}|\tau_a-\tau_b|^{2(1+p_a+p_b)}\times\nn\\
\nn\\
&&\qquad\qquad\qquad\qquad\quad\times\int\prod_{i<j}^{N_f-N_c-1}|u_i-u_j|^2\prod_{i=1}^{N_f-N_c-1}\prod_{a=1}^{N_f}|u_i-\tau_a|^{-2(1+p_a)}\udl{^2u_{N_f-N_c-1}}\, ,\nn\\
\label{onemonopoleDF}
\ee
which corresponds to the $2d$ Coulomb limit of the duality with one monopole in the superpotential proposed in \cite{Benini:2017dud}. Repeated application of the duality yields the following relation between the original integral and the same integral with 
dimension decreased by one unit
\be
\label{Selberg_funk_Relat}
   &&I_N(\alpha_1,\alpha_2,\alpha_3)=\left(\frac{\gamma(-Nb^2)}{\gamma(-b^2)}\right)
   \frac{1}
   {\gamma(2b\alpha_1)\gamma(2b\alpha_2)\gamma(2b\alpha_3+(N-1)b^2)}\;
   \times\nn\\
   &&\qquad\qquad\qquad\qquad\qquad\qquad\qquad\qquad\qquad\qquad\times I_{N-1}(\alpha_1+b/2,\alpha_2+b/2,\alpha_3)\, ,\nn\\
\ee
{with $\gamma(x)=\Gamma(x)/\Gamma(1-x)$.} Iterating this procedure $N$ times, we obtain the evaluation formula
\begin{equation}\label{INev}
   I_N(\alpha_1,\alpha_2,\alpha_3)=
   \prod_{k=1}^N\left(\frac{\gamma(-kb^2)}{\gamma(-b^2)}\right)
   \prod_{j=0}^{N-1}\frac{1}{\gamma(2b\alpha_1+jb^2)\gamma(2b\alpha_2+jb^2)
   \gamma(2b\alpha_3+jb^2)}\, .
\end{equation}

At this point one would like to find an expression which depends parametrically on $N$, so that we can analytically continue it to non-integer values lifting the screening condition \eqref{onshell} and reconstructing the structure constant $C(\alpha_1,\alpha_2,\alpha_3)$ for generic values of momenta. This was done by \cite{Dorn:1994xn,Zamolodchikov:1995aa} (see also \cite{Teschner:2003en})
\begin{equation}\label{3point}
    C(\alpha_1,\alpha_2,\alpha_3)=
    \Bigl[\pi\mu\gamma(b^2)b^{2-2b^2}\Bigr]^{\frac{(Q-\alpha)}{b}}
    \frac{\Upsilon'(0)\prod_{k=1}^3\Upsilon(2\alpha_k)}
    {\Upsilon(\alpha-Q)\prod_{k=1}^3\Upsilon(\alpha-2\alpha_k)},
\end{equation}
where  $\Upsilon(x)$ satisfies the periodicity relations
\begin{equation}
 \begin{aligned}
   &\Upsilon(x+b)=\gamma(bx)b^{1-2bx}\Upsilon(x)\\
   &\Upsilon(x+b^{-1})=\gamma(b^{-1}x)b^{2b^{-1}x-1}\Upsilon(x)\, .
 \end{aligned}
\end{equation}
In particular, this is the unique function having the correct set of zero points
\begin{equation}
    x=\begin{cases}
        -mb-nb^{-1},\\
        Q+mb+nb^{-1}, 
      \end{cases}\;\;
     m,n=0,1,2,\dots\, ,
\end{equation}
which means that the analytic continuation \eqref{3point} of \eqref{INev} truly is the full 3-point function.

We will show that this story has a $3d$ counterpart. The evaluation formula  of the $I_N$ integral \eqref{INev} is actually {\it uplifted} to a genuine $3d$ duality relating a $U(N)$ theory with one adjoint chiral and one fundamental flavor and a Wess-Zumino (WZ) model with $3N$
chiral fields, which has been recently proposed in \cite{Benvenuti:2018bav}.

Here we provide an alternative and very insightful derivation of this duality which retraces the steps that are done in CFT. Basically we derive it by iterating two fundamental dualities, the one with one monopole in the superpotential  \cite{Benini:2017dud} and the Aharony duality  \cite{Aharony:1997gp}.

\begin{figure}[t]
	\centering
	\makebox[\linewidth][c]{
	\includegraphics[scale=0.34]{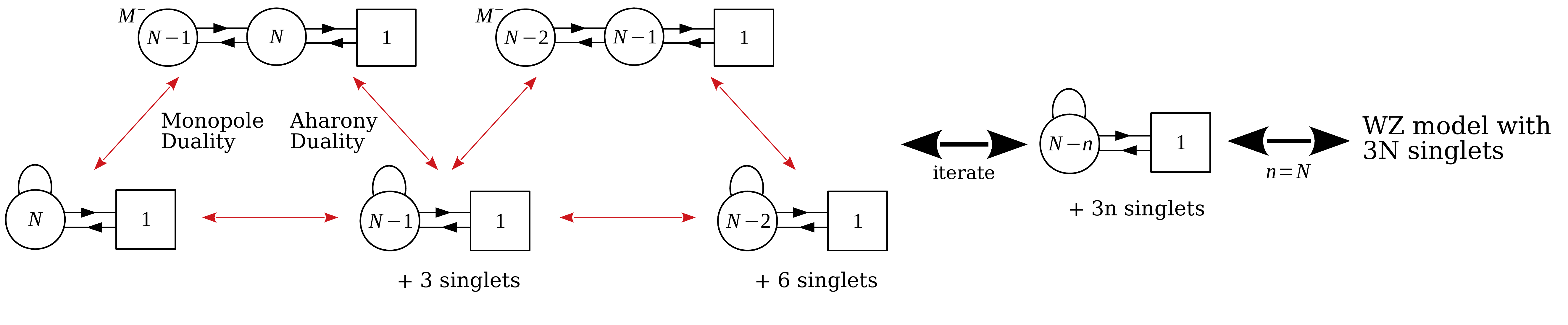}}
	\caption{Diagrammatic representation of the manipulations we perform in the derivation of the duality.
	}
	\label{intro1}
\end{figure}
 
We start considering an auxiliary quiver theory with $U(N-1)\times U(N)$ gauge group and with  single linear monopole superpotential  turned on at the $U(N-1)$ node.
From this theory, we can move in two directions as shown in Figure \ref{intro1}. If we apply first the duality with one monopole superpotential on the $U(N-1)$ node this confines and we return to the original theory. Instead if we apply Aharony duality on the $U(N)$ node, this is confined and we reach a dual frame where we basically have the same theory but with a lower rank and three extra singlets. If we iterate $N$ times this sequential combination of the one-monopole and the Aharony duality, we end up with a WZ model with $3N$ gauge singlets, which is the claimed dual theory.

\begin{figure}[t]
	\centering
	\makebox[\linewidth][c]{
	\includegraphics[scale=0.42]{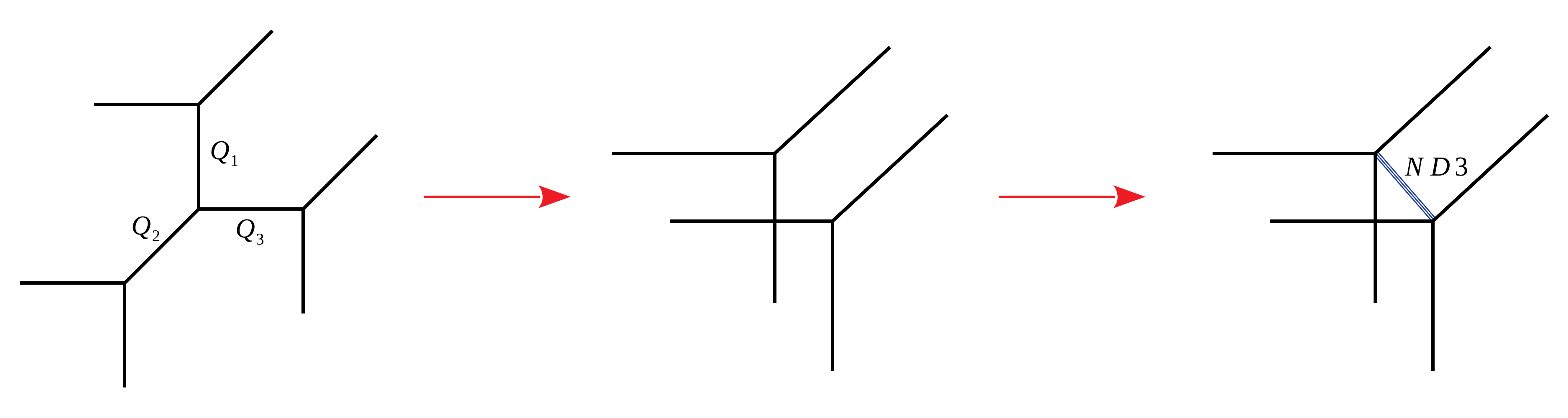}}
	\caption{Schematic representation of the geometric transition from the $5d$ $T_2$ theory to the $3d$ $U(N)$ theory. At the first step we un-resolve the singularity of the quantized $\mathbb{P}^1$. At the second step, we move apart the two sets of intersecting five-branes, between which can then stretch $N$ D3 branes. On this D3 branes lives the $U(N)$ theory. }
	\label{geometrictransition}
\end{figure}

Interestingly, we are also able to make sense of the analytic continuation in the rank $N$ from the gauge theory point of view. Indeed, we can perform an analytic continuation of the $3d$ partition function (for example on $S^3$ or $S^2\times S^1$) of our WZ theory in terms of $5d$ hypermulipelts (on $S^5$ or $S^4\times S^1$) and show that it can be regarded as the residue of the $5d$ partition function of the $T_2$ theory when one of the K\"ahler parameters is quantized. For this quantized value the theory undergoes geometric transition and we can then interpret the $U(N)$ theory as the defect theory on the $N$ stretched D3 branes, as shown in Figure \ref{geometrictransition}.
This interpretation  was first put forward in \cite{Aganagic:2013tta,Aganagic:2014oia} in the context of the Gauge-Liouville triality and here we can see a very neat realization of this idea.

One can profitably follow the philosophy explained in this paper to derive new non-trivial $3d$ dualities from known duality relations for DF integrals. For example in a companion paper \cite{PS} we will uplift the integral identities found in \cite{Fateev:2007qn} for the correlators of 3 arbitrary and $k$ degenerate primaries in Liouville theory. This will lead to a $3d$ duality for a $U(N)$ gauge theory with one adjoint chiral and $k+1$ flavors, which generalizes the one discussed in this paper. 

\section{The \boldmath$3d$ duality}
\label{3d}

\subsection{Statement of the duality and map of the chiral ring generators}

As we mentioned in the Introduction, the  evaluation formula  of the $I_N$ integral \eqref{INev} can be {\it uplifted} to the genuine $3d$ duality recently proposed in \cite{Benvenuti:2018bav}, which can be considered as a non-abelian generalization of the duality between SQED with one flavor and the XYZ model \cite{Aharony:1997bx}. In this section we review this duality and present an alternative derivation.

The duality relates:

\medskip
\noindent \textbf{Theory A}: $U(N)$ gauge theory with one adjoint chiral $\Gp$, one fundamental flavor $P$, $\tilde{P}$, $N$ chiral singlets $\gb_j$ and superpotential
\be
\mathcal{W}=\sum_{j=1}^N\gb_j\Tr{\Gp^j}\, .
\ee

\medskip
\noindent \textbf{Theory B}: WZ model with $3N$ chiral singlets $\ga_j$, $T_j^\pm$ and cubic superpotential
\be
\hat{\mathcal{W}}=\sum_{i,j,l=1}^N\ga_iT_j^+T_{N-l+1}^-\gd_{i+j+l,2N+1}\, .
\ee

\medskip
\noindent A key role is played by the $\beta$-fields, whose equations of motion have the effect of setting to zero all the Casimir operators $\Tr\Gp^j$ in the chiral ring. Indeed, such Casimir operators are expected to violate the unitarity bound, which means that they become a decoupled free sector of the theory in the IR. In \cite{Benvenuti:2017lle} it was shown that the correct way to deal with such operators is to flip them by introducing some additional gauge singlet chiral fields that couple to them in the superpotential. 
The $\beta$-fields have the interesting property of vanishing in the chiral ring because of some quantum effects. One argument for this was already used in \cite{Benvenuti:2017kud}, following \cite{Kutasov:1995np}. If any of these gauge singlets, say $\beta_k$, acquires a non-vanishing VEV, then a superpotential of the form $\mathcal{W}=\Tr\Gp^k$ is generated, but the theory with such a superpotential has no stable supersymmetric vacua because of the very low number of flavors.

The global symmetry group of Theory A consists of two abelian flavor symmetries, one rotating the adjoint chiral and the second rotating the fundamental flavor, and the topological symmetry
\be
U(1)_{\tau}\times U(1)_\mu\times U(1)_\zeta\, .
\label{globalsymm}
\ee
For each of the $U(1)$ flavor symmetries  we can turn on a real mass, which we denote respectively with $\mathrm{Re}(\tau)$ and $\mathrm{Re}(\mu)$, while for the topological symmetry we turn on an FI parameter $\zeta$. The two abelian flavor symmetries can mix with the R-symmetry in the IR, so to each of them we also associate a mixing coefficient. We parametrize the mixing with $U(1)_{\tau}$ by $1-R$ and the one with $U(1)_\mu$ by $r$.
 
The one in \eqref{globalsymm} is also the global symmetry group of Theory B because of the cubic superpotential. In Table \ref{charges} we summarize the charges of all the chiral fields of the two theories under the global symmetries and we also specify our parametrization of the R-charges in terms of $r$ and $R$.

\begin{table}[t]
\centering
\scalebox{1}{
\begin{tabular}{c|ccc|c}
{} & $U(1)_{\tau}$ & $U(1)_\mu$ & $U(1)_\zeta$ & $U(1)_R$ \\ \hline
$P$ & 0 & 1 & 0 & $r$ \\
$\tilde{P}$ & 0 & 1 & 0 & $r$ \\
$\Gp$ & 2 & 0 & 0 & $2(1-R)$ \\
$\gb_j$ & $-2j$ & 0 & 0 & $2-2j(1-R)$ \\ 
$\mathfrak{M}^\pm_{\Gp^s}$ & $-2(N-s-1)$ & -1 & $\pm1$ & $1-r-2(N-s-1)(1-R)$ \\
$\Tr\left(\tilde{P}\Gp^sP\right)$ & $2s$ & 2 & 0 & $2r+2s(1-R)$ \\ \hline
$\ga_i$ & $2(i-1)$ & 2 & 0 & $2r+2(i-1)(1-R)$ \\
$T^+_j$ & $-2(N-j)$ & $-1$ & 1 & $1-r-2(N-j)(1-R)$ \\
$T^-_{N-l+1}$ & $-2(N-l)$ & $-1$ & $-1$ & $1-r-2(N-l)(1-R)$ \\
\end{tabular}}
\caption{Charges under the global symmetries and R-charges of all the chiral fields and all the chiral ring generators of the dual theories.}
\label{charges}
\end{table}

As a first check of the duality, we can map the chiral ring generators of the two theories. This is basically a non-abelian generalization of the map for the SQED/XYZ duality, where the two monopole operators and the meson of the electric theory are mapped into the three gauge singlets of the magnetic theory. In our case, we can dress these fundamental operators with powers of the adjoint chiral $\Gp$ \cite{Cremonesi:2013lqa}. In total, we get $3N$ independent operators that generate the chiral ring of Theory A
\be
\mathfrak{M}^\pm_{\Gp^s},\qquad \Tr\left(\tilde{P}\Gp^sP\right),\qquad s=0,\cdots,N-1\, .
\ee
Their charges under the global symmetries are listed in Table \ref{charges}. These operators directly map under the duality into the $3N$ singlets of the WZ theory
\be
\mathfrak{M}^+_{\Gp^s}\quad&\leftrightarrow&\quad T^+_{s+1} \nn\\
\mathfrak{M}^-_{\Gp^s}\quad&\leftrightarrow&\quad T^-_{N-s}\nn\\
\Tr\left(\tilde{P}\Gp^sP\right)\quad&\leftrightarrow&\quad \ga_{s+1},\qquad s=0,\cdots,N-1\, .
\ee

We can also look at the localized  partition functions on the squashed three-sphere $S^3_b$
\cite{Jafferis:2010un,Hama:2010av,Hama:2011ea}, which we write following the convention of the last two references.
The localized partition function  reduces to a matrix integral over the zero mode of the real scalar in the Cartan of the gauge group. A chiral field of R-charge $R_Q$ and real mass $z$ for a $U(1)_z$ symmetry contributes by $\sbfunc{i\frac{Q}{2}(1-R_Q) -z}$, where the double sine function is defined in Appendix \ref{multuiplesine}. Finally mixed Chern-Simons couplings contribute as exponentials of quadratic forms of real masses for the flavor or R-symmetries and real scalars.

At the level of three-sphere partition functions the duality is encoded in the following integral identity:
\be
\mathcal{Z}_{\mathcal{T}_A}=\prod_{j=1}^{N}\sbfunc{-i\frac{Q}{2}+2j\tau}\int\udl{x_N}\e^{2\pi i\zeta\sum_\ga x_\ga}\frac{\prod_{\ga,\gb=1}^Ns_b\left(i\frac{Q}{2}+(x_\ga-x_\gb)-2\tau\right)}{\prod_{\ga<\gb}^Ns_b\left(i\frac{Q}{2}\pm(x_\ga-x_\gb)\right)}\times&&\nn\\
\qquad\qquad\qquad\qquad\qquad\qquad\qquad\qquad\qquad\qquad\qquad\times\prod_{\ga=1}^Ns_b\left(i\frac{Q}{2}\pm x_\ga-\mu\right)&=&\nn\\
=\prod_{j=1}^{N}\sbfunc{i\frac{Q}{2}-2\mu-2(j-1)\tau}\sbfunc{-\zeta+\mu+2(N-j)\tau}\sbfunc{\zeta+\mu+2(j-1)\tau}&=&\mathcal{Z}_{\mathcal{T}_B}\, .\nn\\
\label{rankstabid}
\ee
The integration measure is defined including the Weyl symmetry factor of the gauge group
\be
\udl{x_{N}}=\frac{1}{N!}\prod_{i=1}^N\udl{x_i}\, .
\ee
The mixing coefficients with the R-symmetry and the real masses for the $U(1)_\tau$ and $U(1)_\mu$ symmetries appear in the localized partition function through the holomorphic combinations \cite{Jafferis:2010un}
\bea
\tau=\mathrm{Re}(\tau)+i\frac{Q}{2} (1-R)\,, \qquad \mu=\mathrm{Re}(\mu) +i\frac{Q}{2} r\,.
\eea
Notice that by looking at the argument of the double-sine functions  we can read out the R-charges of the various fields in \eqref{rankstabid} and see that they are consistent with our parametrization in Table \ref{charges}.

This integral identity already appeared in the mathematical literature in \cite{fokko} and connected to this $3d$ duality in \cite{Aghaei:2017xqe}. 


\subsection{Derivation through stabilization}
\label{piecewise}
Following the logic we sketched in the Introduction, we will now  propose a derivation of the duality that retraces the steps performed in \cite{Fateev:2007qn} to evaluate the DF integral associated to the 3-point function in Liouville theory. This derivation iterates  some basic dualities, namely the Aharony duality \cite{Aharony:1997gp} and some variants with monopole superpotential proposed in \cite{Benini:2017dud} that we are now going to quickly review. In particular, we stress the effect of the contact terms predicted by these dualities, which play a fundamental role in our derivation.

\subsubsection{Basic dualities}\label{basicdualities}

\subsubsection*{Two-monopole duality}

The most fundamental of the dualities we are going to need was first proposed in \cite{Benini:2017dud}:

\medskip
\noindent \textbf{Theory 1}: $U(N_c)$ with $N_f$ fundamental flavors and superpotential
\be
\mathcal{W}=\mathfrak{M}^++\mathfrak{M}^-\, .
\ee

\medskip
\noindent \textbf{Theory 2}: $U(N_f-N_c-2)$ with $N_f$ fundamental flavors, $N_f^2$ singlets (collected in a matrix $M_{ij}$) and superpotential
\be
\hat{\mathcal{W}}=\sum_{i,j=1}^{N_f}M_{ij}\tilde{q}_iq_j+\hat{\mathfrak{M}}^++\hat{\mathfrak{M}}^-\, .
\ee

\medskip
\noindent The monopole superpotential completely breaks both the axial and the topological symmetry, so that the global symmetry group of the two theories is $SU(N_f)\times SU(N_f)$. At the level of three-sphere partition functions, this duality is represented by the following integral identity
\be
\mathcal{Z}_{\mathcal{T}_1}&=&\frac{1}{N_c!}\int\prod_{i=1}^{N_c}\udl{x_i}\frac{\prod_{i=1}^{N_c}\prod_{a=1}^{N_f}s_b\left(i\frac{Q}{2}\pm(x_i+M_a)-\mu_a\right)}{\prod_{i<j}^{N_c}s_b\left(i\frac{Q}{2}\pm(x_i-x_j)\right)}=\nn\\
&=&\frac{1}{(N_f-N_c-2)!}\prod_{a,b=1}^{N_f}s_b\left(i\frac{Q}{2}-(\mu_a+\mu_b-M_a+M_b)\right)\times\nn\\
&\times&\int\prod_{i=1}^{N_f-N_c-2}\udl{x_i}\frac{\prod_{i=1}^{N_f-N_c-2}\prod_{a=1}^{N_f}s_b\left(\pm(x_i-M_a)+\mu_a\right)}{\prod_{i<j}^{N_f-N_c-2}s_b\left(i\frac{Q}{2}\pm(x_i-x_j)\right)}=\mathcal{Z}_{\mathcal{T}_2}\, ,
\label{twomonopoles}
\ee
where $M_a$, $\mu_a$ are real masses corresponding to the Cartan subalgebra of the diagonal and the anti-diagonal combinations of the two $SU(N_f)$ flavor symmetries. The vector masses sum to zero $\sum M_a=0$, while the axial masses have to satisfy the constraint
\be
2\sum_{a=1}^{N_f}\mu_a=iQ(N_f-N_c-1)\, ,
\label{constraintCONF2}
\ee
which is often referred to in the mathematical literature as ``balancing condition". This constraint implies that the two fundamental monopoles are exactly marginal and is consistent with the fact that the axial symmetry is broken. Notice also that we have no FI contribution to the partition function, which is instead consistent with the fact that the topological symmetry is broken. In Sec.~ \ref{2dlimitonemonopole} we will show how this duality, when formulated on $S^2\times S^1$, reduces in the $2d$ Coulomb limit to the basic duality between complex DF integrals used in \cite{Fateev:2007qn}.

As shown in \cite{Benini:2017dud}, from the two-monopole duality we can derive various other dualities by performing suitable real mass deformations.

\subsubsection*{One-monopole duality}

If we start with $N_f+1$ flavors and integrate one of them out, we restore a combination of the axial and the topological symmetry, obtaining  a duality for theories with only one of the two monopoles turned on in the superpotential:

\medskip
\noindent \textbf{Theory 1}: $U(N_c)$ with $N_f$ fundamental flavors and superpotential
\be
\mathcal{W}=\mathfrak{M}^-\, .
\ee

\medskip
\noindent \textbf{Theory 2}: $U(N_f-N_c-1)$ with $N_f$ fundamental flavors, $N_f^2$ singlets (collected in a matrix $M_{ij}$), an extra singlet $S^+$ and superpotential
\be
\hat{\mathcal{W}}=\sum_{i,j=1}^{N_f}M_{ij}\tilde{q}_iq_j+\hat{\mathfrak{M}}^++S^+\hat{\mathfrak{M}}^-\, .
\ee

\medskip
\noindent Implementing the real mass deformation on the partition functions, we get the following identity
\be
\mathcal{Z}_{\mathcal{T}_1}&=&\frac{1}{N_c!}\int\prod_{i=1}^{N_c}\udl{x_i}\e^{i\pi\left(\sum_{i=1}^{N_c}x_i\right)(\eta-iQ)}\frac{\prod_{i=1}^{N_c}\prod_{a=1}^{N_f}s_b\left(i\frac{Q}{2}\pm(x_i+M_a)-\mu_a\right)}{\prod_{i<j}^{N_c}s_b\left(i\frac{Q}{2}\pm(x_j-x_i)\right)}=\nn\\
&=&\frac{1}{(N_f-N_c-1)!}\e^{-i\pi\left(2\sum_{a=1}^{N_f}M_a\mu_a+(\eta-iQ)\sum_{a=1}^{N_f}M_a\right)}s_b\left(i\frac{Q}{2}-\eta\right)\times\nn\\
&&\quad\times\prod_{a,b=1}^{N_f}s_b\left(i\frac{Q}{2}-(\mu_a+\mu_b-M_a+M_b)\right)\times\nn\\
&&\quad\times\int\prod_{i=1}^{N_f-N_c-1}\udl{x_i}\e^{i\pi\eta\sum_{i=1}^{N_c}x_i}\frac{\prod_{i=1}^{N_f-N_c-1}\prod_{a=1}^{N_f}s_b\left(\pm(x_i-M_a)+\mu_a\right)}{\prod_{i<j}^{N_f-N_c-1}s_b\left(i\frac{Q}{2}\pm(x_j-x_i)\right)}=\mathcal{Z}_{\mathcal{T}_2}\, .\nn\\
\label{onemonopole}
\ee
Notice that an FI contribution has been generated, which is consistent with the fact that a combination of the axial and the topological symmetry has been restored. Indeed, we now have one additional parameter $\eta$
 and the balancing condition has become
\be
\eta+2\sum_{a=1}^{N_f}\mu_a=iQ(N_f-N_c)\, .
\label{constraintM-}
\ee

\subsubsection*{Aharony duality}

Starting from the two-monopole duality with $N_f+2$ flavors and integrating two of them out we completely restore both the axial and the topological symmetry and we get Aharony duality \cite{Aharony:1997gp}:

\medskip
\noindent \textbf{Theory 1}: $U(N_c)$ with $N_f$ flavors and superpotential $\mathcal{W}=0$.

\medskip
\noindent \textbf{Theory 2}: $U(N_f-N_c)$ with $N_f$ flavors, $N_f^2$ singlets (collected in a matrix $M_{ij}$), two extra singlets $S^{\pm}$ and superpotential $\hat{\mathcal{W}}=\sum_{i,j=1}^{N_f}M_{ij}\tilde{q}_iq_j+S^-\hat{\mathfrak{M}}^++S^+\hat{\mathfrak{M}}^-$.

\medskip
\noindent At the level of partition functions, the result of the real mass deformation is
\be
\mathcal{Z}_{\mathcal{T}_1}&=&\frac{1}{N_c!}\int\prod_{i=1}^{N_c}\udl{x_i}\e^{i\pi\xi\left(\sum_{i=1}^{N_c}x_i\right)}\frac{\prod_{i=1}^{N_c}\prod_{a=1}^{N_f}s_b\left(i\frac{Q}{2}\pm(x_i+M_a)-\mu_a\right)}{\prod_{i<j}^{N_c}s_b\left(i\frac{Q}{2}\pm(x_j-x_i)\right)}=\nn\\
&=&\e^{-i\pi\xi\sum_{a=1}^{N_f}M_a}s_b\left(i\frac{Q}{2}-\frac{iQ(N_f-N_c+1)-2\sum_{a=1}^{N_f}\mu_a\pm\xi}{2}\right)\times\nn\\
&\times&\prod_{a,b=1}^{N_f}s_b\left(i\frac{Q}{2}-(\mu_a+\mu_b-M_a+M_b)\right)\times\nn\\
&\times&\frac{1}{(N_f-N_c)!}\int\prod_{i=1}^{N_f-N_c}\udl{x_i}\e^{i\pi\xi\sum_{i=1}^{N_c}x_i}\frac{\prod_{i=1}^{N_f-N_c}\prod_{a=1}^{N_f}s_b\left(\pm(x_i-M_a)+\mu_a\right)}{\prod_{i<j}^{N_f-N_c}s_b\left(i\frac{Q}{2}\pm(x_j-x_i)\right)}=\mathcal{Z}_{\mathcal{T}_2}\, .\nn\\
\label{aha}
\ee

An important remark is in order. Notice that the real mass deformation has produced some mixed Chern-Simons couplings between background vector multiplets for the global symmetries, both in the one-monopole duality and in Aharony duality. These are related to contact terms in the two-point functions of the corresponding conserved currents as shown in \cite{Closset:2012vg, Closset:2012vp}, where it was also discussed their role in dualities, in particular their importance in order for the matching of partition functions on $S^3$ or $S^2\times S^1$ to work. In the case of the one-monopole duality, we can see that they represent an interaction between the combination of the axial and the topological symmetry that has been restored and the flavor symmetry. Instead, in the case of Aharony duality we have an interaction between the topological and the flavor symmetry. These mixed background Chern-Simons couplings are encoded in the phases that appear as prefactors in the partition functions of the magnetic duals in \eqref{onemonopole} and \eqref{aha}. Some of these phases would simply  be equal to one when imposing that the vector masses in the Cartan of the non-abelian symmetry $SU(N_f)$ sum to zero $\sum_a M_a=0$. However, when the symmetry is  gauged to become a $U(N_f)$ gauge node, as for example if we  apply the duality  inside a quiver gauge theory, then these contributions become non-trivial  BF couplings.

As we will discuss  in more details in Sec.~ \ref{monopolequantumnumbers}, these BF couplings play a fundamental role, since they modify the charges of the monopoles under the global symmetries. In some cases, they might become uncharged under some of the global symmetries and also exactly marginal because of a BF coupling between the R-symmetry and the gauge symmetry that makes the monopoles have R-charge 2. When this happens, a monopole superpotential may be generated, which breaks the abelian symmetries under which the monopole is still charged.
In other cases, the opposite might happen, that is a monopole superpotential term disappears after applying one of the two dualities on an adjacent node since the monopole quantum numbers have changed because of the BF couplings.

The bottom line is that if we consider applying the duality with one monopole or Aharony duality on a node inside a quiver, it will also affect the adjacent gauge nodes by modifying the charges of the corresponding monopole operators. This remark is fundamental to understand the  derivation of the duality we are going to present.

\subsubsection{Derivation of the duality}

As we sketched in Figure \ref{intro1}, the idea is to combine the one-monopole and the Aharony duality to find a dual frame for Theory A with lower rank and  some extra singlets.

\begin{enumerate}[1)]
\item The first step consists in viewing  Theory A as the result of the application of the one-monopole duality to an auxiliary $U(N-1)\times U(N)$ quiver theory. The quiver has a flavor attached to the $U(N)$ node and $N-1$ singlets $\beta_k$ flipping the traces of the $(k-1)$-th powers
of the meson constructed with the bifundamental chirals $Q$, $\tilde{Q}$ connecting the two gauge nodes
\be
\Tr_N \mathbb{M}^{k-1}\quad k=1,\cdots,N\, ,
\ee
where $\Tr_N$ denotes the trace over the $U(N)$ color indices and
\be
\mathbb{M}=\Tr_{N-1}\tilde{Q}Q
\ee
transforms in the adjoint representation of $U(N)$. In the auxiliary quiver theory there is also the negative fundamental  monopole of the $U(N-1)$ node turned on in the superpotential and a BF coupling between the axial symmetry $U(1)_\tau$ and the gauge symmetry of the $U(N)$ node\footnote{Because of this BF coupling, the fundamental monopoles at the $U(N)$ node with opposite magnetic charge  have different charge under $U(1)_\tau$, which implies that charge conjugation is broken in the auxiliary theory  (see Sec.~ \ref{monopolequantumnumbers})}. This BF coupling compensates a similar BF coupling which is generated when we apply the one-monopole duality to the $U(N-1)$ node, which confines since the number of flavors connected to it is $N$, yielding a $U(N)$ theory with one flavor. Moreover, the matrix of gauge singlets $M$ appearing in the magnetic dual of the one-monopole duality reconstructs exactly the adjoint chiral for the $U(N)$ node, while the singlet $S^+$ is identified with the $\beta_N$ singlet field. So we recovered Theory A.  

\item The second step consists in starting from the auxiliary quiver theory and applying the  Aharony duality to the $U(N)$ node, which confines since the number of flavors connected to it is $N$. Hence, we obtain a $U(N-1)$ theory with one flavor. The matrix of gauge singlets $M$ in the magnetic dual of the Aharony duality then reconstructs  the adjoint chiral for the $U(N-1)$ node,
the fundamental flavor and a singlet, while two more singlets come from $S^\pm$. Moreover, as we reviewed in the previous section, the Aharony duality produces some contact terms that become BF couplings for the  $U(N-1)$ node. These BF couplings have the effect of changing the quantum numbers of the monopole operator, which is removed from the
 superpotential  (see Sec.~ \ref{monopolequantumnumbers}). So we obtain a dual frame for Theory A which is actually the same theory but with rank decreased by one unit and three extra singlets. These three singlets map to the highest dressed monopoles and mesons of the theory with $U(N)$ gauge group. Indeed, in the $U(N-1)$ frame we can only construct $3(N-1)$ dressed monopoles and mesons, which map to the same operators in
the original $U(N)$ theory.
\end{enumerate}

We thus see that the sequential application of the one-monopole and the Aharony duality only decreases the rank of theory A (besides producing extra singlets).  For this reason, we say that Theory A is \emph{stable} under the sequential application of these two basic dualities. 
If we iterate this procedure $N$ times, we completely confine the original gauge node and end up with a WZ model with $3N$ gauge singlets, which is the claimed dual theory.

We can repeat the steps we just described in field theory but at the level of partition functions, thus providing a new analytical proof of the identity \eqref{rankstabid}. This is useful to better understand the role of the contact terms dictated by the basic dualities. We start considering the partition function of Theory A
\be
\mathcal{Z}_N(\tau,\zeta,\mu)\equiv\mathcal{Z}_{\mathcal{T}_A}&=&\prod_{j=1}^N\sbfunc{-i\frac{Q}{2}+2j\tau}\frac{1}{N!}\int\prod_{\ga=1}^N\udl{x_\ga}\e^{2\pi i\zeta\sum_\ga x_\ga}\times\nn\\
&\times&\frac{\prod_{\ga,\gb=1}^Ns_b\left(i\frac{Q}{2}+(x_\ga-x_\gb)-2\tau\right)}{\prod_{\ga<\gb}^Ns_b\left(i\frac{Q}{2}\pm(x_\ga-x_\gb)\right)}\prod_{\ga=1}^Ns_b\left(i\frac{Q}{2}\pm x_\ga-\mu\right)\, .\nn\\
\ee
The first step of the derivation consists of replacing the contribution of the adjoint chiral with an auxiliary $U(N-1)$ integral using \eqref{onemonopole}, where we identify $\Gp$ with the matrix $M$. In this way, we get the partition function of the auxiliary theory
\be
\mathcal{Z}_N(\tau,\zeta,\mu)&=&\prod_{j=1}^{N-1}\sbfunc{-i\frac{Q}{2}+2j\tau}\frac{1}{(N-1)!}\int\prod_{\ga'=1}^{N-1}\udl{y_{\ga'}}\frac{\e^{-2\pi iN\tau\sum_{\ga'}y_{\ga'}}}{\prod_{\ga'<\gb'}^{N-1}s_b\left(i\frac{Q}{2}\pm(y_{\ga'}-y_{\gb'})\right)}\times\nn\\
&\times&\frac{1}{N!}\int\prod_{\ga=1}^N\udl{x_\ga}\frac{\e^{2\pi i(\zeta-(N-1)\tau)\sum_\ga x_\ga}}{\prod_{\ga<\gb}^Ns_b\left(i\frac{Q}{2}\pm(x_\ga-x_\gb)\right)}\prod_{\ga=1}^Ns_b\left(i\frac{Q}{2}\pm x_\ga-\mu\right)\times\nn\\
&\times&\prod_{\ga'=1}^{N-1}s_b\left(i\frac{Q}{2}\pm(x_\ga+y_\ga')-\tau\right)\, .\nn\\
\label{adjaha1flavconf1}
\ee
Notice the shift in the FI parameter of the $U(N)$ node by an amount proportional to the axial mass $\tau$, which represents the BF coupling between the $U(1)_\tau$ symmetry and the gauge symmetry we mentioned above.

Now we can remove the original integral by means of Aharony duality \eqref{aha}, which in the case $N_f=N_c=N$ becomes an evaluation formula
\be
\mathcal{Z}_N(\tau,\zeta,\mu)&=&\prod_{j=1}^{N-1}\sbfunc{-i\frac{Q}{2}+2j\tau}s_b\left(i\frac{Q}{2}-2\mu\right)s_b\left(\zeta+\mu\right)\times\nn\\
&\times& s_b\left(-\zeta+\mu+2(N-1)\tau\right)\frac{1}{(N-1)!}\int\prod_{\ga=1}^{N-1}\udl{y_\ga}\e^{-2\pi i(\zeta+\tau)\sum_\ga y_\ga}\times\nn\\
&\times&\frac{\prod_{\ga,\gb=1}^{N-1}s_b\left(i\frac{Q}{2}\pm(y_\ga-y_\gb)-2\tau\right)}{\prod_{\ga<\gb}^{N-1}s_b\left(i\frac{Q}{2}\pm(y_\ga-y_\gb)\right)}\prod_{\ga=1}^{N-1}s_b\left(i\frac{Q}{2}\pm y_\ga-\mu-\tau\right)\, .\nn\\
\label{adjaha1flavconf0}
\ee
At this point, we notice that we have reconstructed the same structure of the original partition function (up to the change of variables $y_i\rightarrow-y_i$), but with a lower dimensional integral, a shift of the parameters and three singlets:
\be
&&\mathcal{Z}_N(\tau,\zeta,\mu)=s_b\left(i\frac{Q}{2}-2\mu\right)s_b\left(\zeta+\mu\right)s_b\left(-\zeta+\mu+2(N-1)\tau\right) \mathcal{Z}_{N-1}(\tau,\zeta+\tau,\mu+\tau)\, .\nn\\
\ee
We refer to this property of the original partition function saying that it is \emph{stabilized} with respect to the two moves we performed. 

Notice that when we applied the Aharony duality to arrive at \eqref{adjaha1flavconf0},
the contact term from \eqref{aha} given by  $\e^{-2\pi i(\zeta-(N-1)\tau)\sum_{\ga}y_\ga}$ becomes a BF coupling for the remaining $U(N-1)$ node.
More precisely, this contains a mixed CS term between the gauge and  the topological symmetry $U(1)_\zeta$. Thus, we conclude that the topological symmetry $U(1)_\zeta$ is not broken and hence the monopole can't be turned on in the superpotential anymore. We also see that $\zeta$ is shifted by $\tau$ and, as we will explain later in Sec.~ \ref{monopolequantumnumbers}, we can interpret this as the fact that the monopoles with positive and negative magnetic charge have different charge under the axial symmetry $U(1)_\tau$, which means that at this stage charge conjugation is broken.

Finally we can use the stabilization property to highly simplify our expression by reducing the dimension of the integral, which from the field theory perspective is the rank of the gauge group. Indeed, iterating $n$ times the two steps that we performed, we get
\be
\mathcal{Z}_N(\tau,\zeta,\mu)&=&\prod_{j=1}^{n}\sbfunc{i\frac{Q}{2}-2\mu-2(j-1)\tau}\sbfunc{-\zeta+\mu+2(N-j)\tau}\times\nn\\
&\times&\sbfunc{\zeta+\mu+2(j-1)\tau}\mathcal{Z}_{N-n}(\tau,\zeta+n\tau,\mu+n\tau)\, .
\ee
If we set $n=N$, we completely confine the original gauge node and obtain the partition function of the WZ dual theory
\be
\mathcal{Z}_{\mathcal{T}_A}&=&\mathcal{Z}_N(\tau,\zeta,\mu)=\prod_{j=1}^N\sbfunc{i\frac{Q}{2}-2\mu-2(j-1)\tau}\times\nn\\
&&\qquad\qquad\quad\times\sbfunc{-\zeta+\mu+2(N-j)\tau}\sbfunc{\zeta+\mu+2(j-1)\tau}=\mathcal{Z}_{\mathcal{T}_B}\, .\nn\\
\ee

\subsection{Superconformal Index}
\label{sci}

We mentioned several times that, when we apply a duality to a given node inside a quiver, contact terms become BF couplings which modify the charges of the monopole operators. The best way to understand this is by looking at the superconformal index, as we show in this section.

Since our discussion focuses on  BF couplings, it is enough to consider the unrefined version of the index where we don't turn on magnetic fluxes for the global symmetries, however by doing so we will not be able to see the contribution of the contact terms.

\subsubsection{Definition and computation}

The unrefined version of the superconformal index is defined as
\be
\mathcal{I}(x,t)=\Tr\left[(-1)^F\e^{-\gb'(E-R-j)}x^{E+j}\prod_i t_i^{F_i}\right]\, ,
\label{indexdef}
\ee
where $E$, $j$ and $R$ are quantum numbers corresponding to the bosonic part of the superconformal group $SO(2,3)\times SO(2)$ for three-dimensional $\mathcal{N}=2$ theories compactified on $S^2$, namely the energy, the R-charge and the third component of the angular momentum, while $t_i$ and $F_i$ are fugacities and charges for global symmetries. Only BPS states which saturate the bound
\be
\acom{\mathcal{Q}}{\mathcal{Q}^\dagger}=E-R-j\ge 0\, ,
\ee
where $\mathcal{Q}$ is one of the four supercharges, contribute to the index, so that this is actually independent from $\gb'$. The superconformal index has been computed as the partition function on $S^2\times S^1$ using supersymmetric localization in \cite{Imamura:2011su}
\be
\mathcal{I}(x,t)=\frac{1}{|\mathcal{W}|}\sum_m\int\frac{\udl{u}}{2\pi i\, u}Z_{cl}Z_{vec}Z_{chir}\, ,
\label{index}
\ee
where $(u,m)$ are the fugacity and the flux for the gauge symmetry and $|\mathcal{W}|$ is the dimension of the Weyl group.

The classical contribution consists of  Chern-Simons terms.
A pure  CS interaction takes the form
\be
Z_{cl}=\prod_{i=1}^{N_c}(-u_i)^{-km_i}\, ,
\ee
where $N_c$ is the rank of the gauge group and $k$ is the level of the CS interaction. A mixed CS coupling  between the gauge vector multiplet and the background vector multiplet for the topological symmetry  (FI coupling) for which  we turn on the fugacity  $\omega$ contributes as
\be
Z_{cl}=\prod_{i=1}^{N_c}\omega^{m_i}\, .
\ee

The contribution of the vector multiplet instead is
\be
Z_{vec}=\prod_{\ga\in\mathfrak{g}}x^{-|\ga(m)|/2}\left(1-u^\ga x^{|\ga(m)|}\right)\, ,
\ee
where $\ga$ are the roots of the lie algebra $\mathfrak{g}$ of the gauge group. 

Finally, we  have contributions from chiral multiplets which transform in the representation $\mathcal{R}$ of the gauge group and $\mathcal{R}_G$ of the global symmetry group
\be
Z_{chi}=\prod_{\gr\in \mathcal{R}}\prod_{\gs\in \mathcal{R}_G}\left((-u)^\gr t^\gs x^{R_Q-1}\right)^{-|\gr(m)|/2}\frac{\xfac{u^{-\gr}t^{-\gs}x^{2-R_Q+|\gr(m)|}}}{\xfac{u^{\gr}t^{\gs}x^{R_Q+|\gr(m)|}}}\, ,
\ee
where $\gr$ and $\gs$ are the weights of the representations $\mathcal{R}$ and $\mathcal{R}_G$,
 $t$ denotes the fugacity for the  global symmetry and $x$ the fugacity for the R-symmetry  and we allowed the possibility of an anomalous R-charge $R_Q$. 

It is worth pointing out an interesting feature of the superconformal index. Once we collect all the contributions and focus on the total prefactor multiplying the $q$-factorial, we can see that each  fugacity $z$ for an abelian symmetry $U(1)_z$ appears  raised to a power $z^{\epsilon_z(\vec m)}$ and $\epsilon_z(\vec m)$ gives the charge of the monopole of magnetic charge $\vec m$ under $U(1)_z$.

As an example, let us consider a $U(N_c)$ gauge theory with $N_f$ fundamental flavors and $\mathcal{W}=0$. The global symmetry group is
\be
SU(N_f)\times SU(N_f)\times U(1)_J\times U(1)_A\, .
\ee
We turn on fugacities $t_a$ and $s_a$ for the diagonal and anti-diagonal combinations of the non-abelian symmetries respectively, $\omega$ for the topological symmetry, $s$ for the axial symmetry and $x$ for the R-symmetry. These fugacities are subjected to the constraint
\be
\prod_{a=1}^{N_f}t_a=1,\qquad\prod_{a=1}^{N_f}s_a=s\, .
\label{onshellaha}
\ee
As in the case of the $S^3$ partition functions, we turn on also a mixing coefficient which forms the {\it holomorphic} combinations $s_a x^r$, since in this case the axial symmetry $U(1)_A$ is the only abelian symmetry that can mix with the R-symmetry and $r$ represents the mixing parameter. Then, the superconformal index reads
\be
\mathcal{I}&=&\sum_{\vec{m}\in\mathbb{Z}^{N_c}}\frac{\prod_{i=1}^{N_c}\omega^{m_i}}{N_c!}\oint\prod_{i=1}^{N_c}\frac{\udl{u_i}}{2\pi i\,u_i}\prod_{i<j}^{N_c}x^{-|m_i-m_j|}\left(1-\frac{u_i}{u_j} x^{|m_i-m_j|}\right)\left(1-\frac{u_j}{u_i} x^{|m_i-m_j|}\right)\times\nn\\
&\times&\prod_{i=1}^{N_c}\prod_{a=1}^{N_f}\left(s_a\,x^{r-1}\right)^{-|m_i|}\frac{\xfac{u_i^{\mp1}t_a^{\mp1}s_a^{-1}x^{2-r+|m_i|}}}{\xfac{u_i^{\pm1}t_a^{\pm1}s_a\,x^{r+|m_i|}}}\, ,
\label{sciaha}
\ee
If we collect all the powers of $x$, we find an overall factor with exponent
\be
\epsilon_x (\vec{m})=-\sum_{i<j}^{N_c}|m_i-m_j|+N_f(1-r)\sum_{i=1}^{N_c}|m_i|\, ,
\ee
which reproduces the R-charge of the monopoles. Instead, the exponent of $s$ gives the charge under the axial symmetry $U(1)_A$
\be
\epsilon_s (\vec{m})=-\sum_{i=1}^{N_c}|m_i|\, .
\ee

In order to test dualities, we will compute the index as a power series in the R-symmetry fugacity. When we expand the integrand of \eqref{index} in powers of $x$, the coefficients of the expansion are monomial in the gauge fugacities $u$. Thus, we can easily evaluate the integral, whose measure $\frac{\udl{u}}{2\pi iu}$ simply selects the constant part in $u$ of these coefficients. In principle, we still have to perform the infinite sum over the magnetic fluxes. Nevertheless, if we are interested in computing the index up to some finite order $h$ in $x$, only those configurations of the magnetic fluxes such that $\epsilon(\vec{m})\le h$ will contribute and these will be finite in number provided that
\be
\epsilon_x(\vec{m})> 0 \qquad\forall \vec{m}\in\mathbb{Z}^{N_c}\, .
\label{epsilonconstraint}
\ee
Since $\epsilon_x(\vec m)$ computes the R-charge of the monopoles, it is  sufficient to check that this condition is true for the fundamental monopole operators, namely those configurations where one of the magnetic fluxes is $\pm1$ and all the others are zero. In the case of $\mathcal{N}=4$ supersymmetry, this condition is satisfied by good theories \cite{Gaiotto:2008ak}.

For $\mathcal{N}=2$ supersymmetry,  the R-charges of the fields depend on various mixing parameters  and in computing the index we should fix them to  values consistent with the condition \eqref{epsilonconstraint}.  Moreover, in order for the $x$-expansion of the integrand to be well-defined we also need to impose that the R-charge of all the chiral fields is contained between 0 and 2. In conclusion, for computing the superconformal index as a power series in $x$, we have to choose the R-symmetry mixing parameters such that
\begin{enumerate}[$\bullet$]
\item the R-charge of the fundamental monopoles is positive;
\item the R-charge $R_Q$ of any chiral field in the theory is $0<R_Q<2$.
\end{enumerate}

\subsubsection{Monopole quantum numbers and BF couplings}
\label{monopolequantumnumbers}

We can use the superconformal index to explicitly see  how turning on  a BF coupling between a background vector field for  a global symmetry and the gauge field modifies the charge of the monopoles under such a symmetry. A BF term contributes to the superconformal index as
\be
\prod_{i=1}^{N_c}z^{\ga m_i}\, ,
\ee
where  $z$ denotes the fugacity for the global abelian symmetry and $\ga$ is the coupling of the BF interaction. Since the overall power of $z$ computes the charge of the monopoles under the $U(1)_z$ symmetry, we see that the effect of introducing such a BF coupling is to shift the charge by the amount $\ga \sum_im_i$. Moreover, since the shift depends on  $m_i$ rather than on $|m_i|$, monopole operators with opposite magnetic fluxes, such as $(\pm1,0,\cdots,0)$, will acquire different charge. This means that by turning on the BF coupling we also break charge conjugation symmetry.

As an example, consider again the $U(N_c)$ theory with $N_f$ flavors and $\mathcal{W}=0$ and imagine to turn on a BF interaction between the diagonal combination of the non-abelian symmetries and the gauge symmetry with coupling $\ga$ 
\be 
\prod_{i=1}^{N_c}\prod_{a=1}^{N_f}s_a^{\ga m_i}=\prod_{i=1}^{N_c}s^{\ga m_i}\, .
\ee
Consequently, the charge of the monopoles under the axial symmetry has changed to
\be
\epsilon_s (\vec{m})=-\sum_{i=1}^{N_c}|m_i|+\ga\sum_{i=1}^{N_c}m_i\, .
\ee
We can also turn on a similar BF coupling with the R-symmetry
\be 
\prod_{i=1}^{N_c}\prod_{a=1}^{N_f}x^{r\ga m_i}=\prod_{i=1}^{N_c}x^{N_f r\ga m_i}\, ,
\ee
which instead has the effect of modifying the R-charge of the monopoles
\be
\epsilon_x (\vec{m})=-\sum_{i<j}^{N_c}|m_i-m_j|+N_f(1-r)\sum_{i=1}^{N_c}|m_i|+N_fr\ga\sum_{i=1}^{N_c}m_i\, .
\ee
Notice also that in this way we are able to preserve the holomorphicity in $s\,x^{N_fr}$, since the total BF term we added is
\be 
\prod_{i=1}^{N_c}\left(\prod_{a=1}^{N_f}s_ax^r\right)^{\ga m_i}=\prod_{i=1}^{N_c}\left(s\,x^{N_f r}\right)^{\ga m_i}\, .
\ee

Once we have understood how BF couplings affect the monopole quantum numbers, we can write the superconformal index of a theory with monopole superpotential. Indeed, we can make a monopole operator exactly marginal, namely having R-charge 2, and uncharged under some of the abelian global symmetries by adding suitable BF couplings. Then a monopole superpotential may be generated, which also breaks the symmetries under which the marginal monopole is still charged.

For example, let's see  how we can turn on a single monopole operator starting from the $U(N_c)$ theory with $N_f$ flavors and no superpotential. The total BF factor in \eqref{sciaha} is
\be
\omega^{\sum_{i=1}^{N_c}m_i}s^{-\sum_{i=1}^{N_c}|m_i|}x^{-\sum_{i<j}^{N_c}|m_i-m_j|+N_f(1-r)\sum_{i=1}^{N_c}|m_i|}\, .
\label{prefactoratsomepoint}
\ee
We can rewrite it in terms of fugacities $z_D$ and $z_A$ for the diagonal and anti-diagonal combinations of the axial and the topological symmetry respectively as
\be
z_D^{\sum_{i=1}^{N_c}(m_i-|m_i|)}z_A^{-\sum_{i=1}^{N_c}(m_i+|m_i|)}x^{-\sum_{i<j}^{N_c}|m_i-m_j|+N_f(1-r)\sum_{i=1}^{N_c}|m_i|}\, ,
\ee
with
\be
z_D=(s\,\omega)^{1/2},\qquad z_A=(s\,\omega^{-1})^{1/2}\, .
\ee
We can see that the positive monopole $\mathfrak{M}^+$ is charged under the anti-diagonal combination of the axial and the topological symmetry, while the negative monopole $\mathfrak{M}^-$ is charged under the diagonal one. 

Now suppose we add a BF coupling that makes the negatively charged monopole exactly marginal, that is we add
\be
x^{(N_f(1-r)-(N_c+1)) \sum_{i=1}^{N_c} m_i}\, ,
\ee
so that the prefactor becomes
\be
z_D^{\sum_{i=1}^{N_c}(m_i-|m_i|)}z_A^{-\sum_{i=1}^{N_c}(m_i+|m_i|)}x^{-\sum_{i<j}^{N_c}|m_i-m_j|+N_f(1-r) \sum_{i=1}^{N_c}(|m_i|+ m_i)-(N_c+1) \sum_{i=1}^{N_c} m_i  }\, ,\nn\\
\ee
We see that the negative monopole is now marginal and can enter the superpotential. This has also the effect of breaking the diagonal combination of $U(1)_A$ and $U(1)_J$, so we correspondingly set $z_D=1$. The constraint on the axial masses $s_a$ \eqref{onshellaha} then reads
\be
\prod_{a=1}^{N_f}s_a=z_A\, .
\label{onemonopolefugacityconstraint}
\ee
The overall prefactor has become
\be
(z_A x^{N_f r})^{-\sum_{i=1}^{N_c}(m_i+|m_i|)}x^{-\sum_{i<j}^{N_c}|m_i-m_j|+N_f \sum_{i=1}^{N_c}(|m_i|+ m_i)-(N_c+1) \sum_{i=1}^{N_c} m_i  }\, ,
\ee
where we can notice the holomorphic dependence $z_A x^{N_f r}$, since now the remaining abelian symmetry  corresponding to the anti-diagonal combination of $U(1)_A$ and $U(1)_J$ can mix with the R-symmetry.

As a consistency check of our prescription for writing the superconformal index of a theory with monopole superpotential, we can compute it for the two theories involved in the one-monople duality discussed in Sec.~ \ref{basicdualities} and check that they actually match. For simplicity, we choose to solve the constraint \eqref{onemonopolefugacityconstraint} with the particular choice
\be
t_a=1,\qquad s_a=z_A^{1/N_f},\qquad a=1,\cdots,N_f\, .
\ee
The superconformal index of the electric theory is obtained as explained above by adding suitable BF couplings which make the negatively charged monopole exactly marginal
\be
\mathcal{I}_{\mathcal{T}_1}&=&\frac{1}{N_c!}\sum_{\vec{m}\in\mathbb{Z}^{N_c}}z_A^{-\sum_{i=1}^{N_c}(m_i+|m_i|)}x^{-\sum_{i<j}^{N_c}|m_i-m_j|+N_f(1-r)\sum_{i=1}^{N_c}(m_i+|m_i|)-(N_c+1)\sum_{i=1}^{N_c}m_i}\times\nn\\
&\times&\oint\prod_{i=1}^{N_c}\frac{\udl{u_i}}{2\pi i\,u_i}\prod_{i<j}^{N_c}\left(1-\left(\frac{u_i}{u_j}\right)^{\pm1} x^{|m_i-m_j|}\right)\prod_{i=1}^{N_c}\prod_{a=1}^{N_f}\frac{\xfac{u_i^{\mp1}z_A^{-1/N_f}x^{2-r+|m_i|}}}{\xfac{u_i^{\pm1}z_A^{1/N_f}x^{r+|m_i|}}}\, .\nn\\
\label{onemonopoleindexA}
\ee
In the magnetic theory, we need to turn on a different BF coupling in order for the positive fundamental monopole to be exactly marginal
\be
\mathcal{I}_{\mathcal{T}_2}&=&\frac{1}{(N_f-N_c-1)!}\frac{\xfac{z_A^2x^{2N_fr-2(N_f-N_c-1)}}}{\xfac{z_A^{-2}x^{2(N_f-N_c)-2N_fr}}}\prod_{a,b=1}^{N_f}\frac{\xfac{z_A^{-2/N_f}x^{2(1-r)}}}{\xfac{z_A^{2/N_f}x^{2r}}}\times\nn\\
&\times&\sum_{\vec{m}\in\mathbb{Z}^{N_f-N_c-1}}z_A^{\sum_{i=1}^{N_f-N_c-1}(-m_i+|m_i|)}x^{N_fr\sum_{i=1}^{N_f-N_c-1}(-m_i+|m_i|)-(N_c-N_f)\sum_{i=1}^{N_f-N_c-1}m_i}\times\nn\\
&\times&x^{-\sum_{i<j}^{N_f-N_c-1}|m_i-m_j|}\oint\prod_{i=1}^{N_f-N_c-1}\frac{\udl{u_i}}{2\pi i\,u_i}\prod_{i<j}^{N_f-N_c-1}\left(1-\left(\frac{u_i}{u_j}\right)^{\pm1} x^{|m_i-m_j|}\right)\times\nn\\
&\times&\prod_{i=1}^{N_f-N_c-1}\frac{\xfac{u_i^{\mp1}z_A^{1/N_f}x^{1+r+|m_i|}}}{\xfac{u_i^{\pm1}z_A^{-1/N_f}x^{1-r+|m_i|}}}\, .
\label{onemonopoleindexB}
\ee
From the exponent of the overall factor of $x$, we see indeed that the R-charge of the monopoles is
\be
\epsilon_{x}(\vec{m})=-\sum_{i<j}^{N_f-N_c-1}|m_i-m_j|+N_f r\sum_{i=1}^{N_f-N_c-1}(-m_i+|m_i|)-(N_c-N_f)\sum_{i=1}^{N_f-N_c-1}m_i\nn\\
\ee
and a configuration of magnetic fluxes of the form $(+1,0,\cdots,0)$ corresponds to a monopole with R-charge 2. Similarly, we can easily verify that the positive monopole is uncharged under the $U(1)_A-U(1)_J$ symmetry.

%
%

As we explained in the previous section, for computing the index of the two theories as a power series in $x$ we need to choose $r$ such that the monopoles have positive R-charge and that all the chiral fields have R-charge between 0 and 2. This gives the constraint
\be
\frac{N_f-N_c-1}{N_f}<r<\frac{N_f-N_c}{N_f}\, .
\ee
We computed the indices of the dual theories for different values of $N_c$ and $N_f$ (see Table \ref{testM-}) and found a perfect match between the coefficients of the two expansions.
\begin{table}[t]
\centering
\scalebox{0.95}{
\setlength{\extrarowheight}{2pt}
\begin{tabular}{c|c|c|c|c}
$N_c$ & $N_f$ & $r$ & $h$ & $\mathcal{I}$ \\ \hline
1 & 3 & $\frac{1}{2}$ & 20 & $1+9 z_A^{2/3}x+\frac{1}{z_A^2}x+36 z_A^{4/3}x^2+\frac{1}{z_A^4}x^2-17x^2+\cdots$ \\
1 & 4 & $\frac{2}{3}$ & 12 & $1+\frac{1}{z_A^2}x^{2/3}+\frac{1}{z_A^4}x^{4/3}+16 \sqrt{z_A}x^{4/3}+\frac{1}{z_A^6}x^2-31x^2+\cdots$ \\
2 & 4 & $\frac{1}{3}$ & 14 & $1+16 \sqrt{z_A}x^{2/3}+\frac{1}{z_A^2}x^{4/3} + 136 z_Ax^{4/3}+800 z_A^{3/2}x^2+\frac{16}{z_A^{3/2}}x^2-31x^2+\cdots$ \\
2 & 5 & $\frac{1}{2}$ & 14 & $1+25 z_A^{2/5}x+\frac{1}{z_A^2}x+325 z_A^{4/5}x^2+\frac{25}{z_A^{8/5}}x^2+\frac{1}{z_A^4}x^2-49x^2+\cdots$ \\
3 & 5 & $\frac{3}{10}$ & 10 & $1+25 z_A^{2/5}x^{3/5}+\frac{1}{z_A^2}x+325 z_A^{4/5}x^{6/5}+\frac{25}{z_A^{8/5}}x^{8/5}+2925 z_A^{6/5}x^{9/5}+\cdots$ \\
3 & 6 & $\frac{5}{12}$ & 12 & $1+36 \sqrt[3]{z_A}x^{5/6}+\frac{1}{z_A^2}x+666 z_A^{2/3}x^{5/3}+\frac{36}{z_A^{5/3}}x^{11/6}+\frac{1}{z_A^4}x^2-71x^2+\cdots$ \\
4 & 6 & $\frac{1}{4}$ & 6 & $1+36 \sqrt[3]{z_A}x^{1/2}+666 z_A^{2/3}x+\frac{1}{z_A^2}x+\frac{36}{z_A^{5/3}}x^{3/2}+8436 z_Ax^{3/2}+\cdots$
\end{tabular}}
\caption{Computation of the superconformal index for different values of $N_c$ and $N_f$ up to order $\mathcal{O}\left(x^h\right)$. In the last column we report the first terms of the expansion.}
\label{testM-}
\end{table}

We can also turn on both the negatively and positively charged fundamental monopoles in the superpotential. Looking at the overall BF coupling \eqref{prefactoratsomepoint} of the theory with $\mathcal{W}=0$, we see that in order to make both the monopoles exactly marginal we can simply set the R-charge of the chirals to the value $r=\frac{N_f-N_c-1}{N_f}$. In this case both the topological and the axial symmetries are broken, so we have to set $\omega=s=1$.

%
%

\subsubsection{Test of the duality between the $U(N)$ theory and the WZ model}

We can also use the superconformal index to test the duality between the $U(N)$ theory and the WZ model by computing it perturbatively in the R-symmetry fugacity $x$ for the two theories and checking that the expansions match order by order. For this purpose, we turn on fugacities $\omega$ for the topological symmetry $U(1)_\zeta$ and $s$, $p$ for the $U(1)_\tau\times U(1)_\mu$ axial symmetries. The mixing of these symmetries with the R-symmetry is parametrized by $1-R$ and $r$ respectively. The superconformal index of Theory A then takes the form
\be
\mathcal{I}_{\mathcal{T}_A}&=&\prod_{j=1}^N\frac{\xfac{s^{2j}x^{2j(1-R)}}}{\xfac{s^{-2j}x^{2-2j(1-R)}}}\sum_{\vec{m}\in\mathbb{Z}^N}\frac{\prod_{i=1}^N\omega^{m_i}}{N!}\oint\prod_{i=1}^N\frac{\udl{u_i}}{2\pi i\,u_i}s^{-2\sum_{i<j}^N|m_i-m_j|}\times\nn\\
&\times& p^{-\sum_{i=1}^N|m_i|}x^{2(R-1)\sum_{i<j}^N|m_i-m_j|-(r-1)\sum_{i=1}^N|m_i|}\prod_{i<j}^N\left(1-\left(\frac{u_i}{u_j}\right)^{\pm1}x^{|m_i-m_j|}\right)\times\nn\\
&\times&\prod_{i,j=1}^N\frac{\xfac{\frac{u_i}{u_j}s^{-2}x^{2R+|m_i-m_j|}}}{\xfac{\frac{u_j}{u_i}s^2x^{2(1-R)+|m_i-m_j|}}}\prod_{i=1}^N\frac{u_i^{\pm1}p^{-1}x^{2-r+|m_i|}}{u_i^{\mp1}px^{r+|m_i|}}\, ,
\label{3ddualityindexA}
\ee
while the index of Theory B is
\be
\mathcal{I}_{\mathcal{T}_B}&=&\prod_{j=1}^N\frac{\xfac{s^{-2(j-1)}p^{-2}x^{2-2(j-1)(1-R)-2r}}}{\xfac{s^{2(j-1)}p^2x^{2(j-1)(1-R)+2r}}}\frac{\xfac{s^{2(N-j)}p\,\omega^{-1}x^{1+2(N-j)(1-R)+r}}}{\xfac{s^{-2(N-j)}p^{-1}\omega\,x^{1-2(N-j)(1-R)-r}}}\times\nn\\
&\times&\frac{\xfac{s^{2(j-1)}p\,\omega\,x^{1+2(j-1)(1-R)+r}}}{\xfac{s^{-2(j-1)}p^{-1}\omega^{-1}x^{1-2(j-1)(1-R-r)}}}\, .
\label{3ddualityindexB}
\ee
In order for the two indices to both have a well-defined expansion in $x$, we need to choose the R-symmetry parameters such that
\be
\frac{2N-3}{2(N-1)}<R<1,\qquad 0<r<2(N-1)R+3-2N\, .
\ee
We verified the matching of the superconformal indices for $N=1,2,3,4$. In Table \ref{sci1flav} we summarize the results of our computations. 

\begin{table}[t]
\centering
\scalebox{0.93}{
\setlength{\extrarowheight}{2pt}
\makebox[\linewidth][c]{
\begin{tabular}{c|c|c|c|c}
$N$ & $R$ & $r$ & $h$ & $\mathcal{I}$ \\ \hline
1 & $\frac{1}{2}$ & $\frac{1}{2}$ & 20 & $1+\frac{\omega}{p}x^{1/2}+\frac{1}{p \omega}x^{1/2}+\frac{\omega^2}{p^2}x+\frac{1}{p^2 \omega^2}x+p^2x+\frac{\omega^3}{p^3}x^{3/2}+\frac{1}{p^3 \omega^3}x^{3/2}+\cdots$ \\
2 & $\frac{3}{4}$ & $\frac{1}{4}$ & 8 & $1+\frac{\omega}{p s^2}x^{1/4}+\frac{1}{p s^2 \omega}x^{1/4}+\frac{\omega^2}{p^2 s^4}x^{1/2}+\frac{1}{p^2 s^4 \omega^2}x^{1/2}+\frac{1}{p^2 s^4}x^{1/2}+p^2 x^{1/2}+\cdots$ \\
3 & $\frac{5}{6}$ & $\frac{1}{6}$ & 4 & $1+\frac{\omega}{p s^4}x^{1/6}+\frac{1}{p s^4 \omega}x^{1/6}+\frac{\omega^2}{p^2 s^8}x^{1/3}+\frac{1}{p^2 s^8 \omega^2}x^{1/3}+\frac{1}{p^2 s^8}x^{1/3}+p^2 x^{1/3}+\cdots$ \\
4 & $\frac{7}{8}$ & $\frac{1}{8}$ & 2 & $1+\frac{\omega}{p s^6}x^{1/8}+\frac{1}{p s^6 \omega}x^{1/8}+\frac{\omega^2}{p^2 s^{12}}x^{1/4}+\frac{1}{p^2 s^{12} \omega^2}x^{1/4}+\frac{1}{p^2 s^{12}}x^{1/4}+p^2 x^{1/4}+\cdots$
\end{tabular}}}
\caption{Computation of the superconformal index for different values of $N$ up to order $\mathcal{O}\left(x^h\right)$. In the last column we report the first terms of the expansion.}
\label{sci1flav}
\end{table}

\section{\boldmath$2d$ Coulomb limit of the superconformal index}

In this section, we explain how to perform the dimensional reduction of the superconformal index that leads to the complex DF integrals. 
For this purpose, we use the identity \cite{Dimofte:2011py}
 \be
 \label{misty}
(-x)^{\frac{|m|-m}{2}}\zeta^{-\frac{|m|-m}{2}}\frac{\xfac{\zeta^{-1}x^{2+r+|m|}}}{\xfac{\zeta\,x^{r+|m|}}}=\frac{\xfac{\zeta^{-1}x^{2+r+m}}}{\xfac{\zeta\,x^{r+m}}}\, ,
\ee
to rewrite the index of our theory\footnote{Notice that we are not changing the definition of the index, which is still defined as in eq.~ \eqref{indexdef}.} in a form which does not contain absolute values of the magnetic fluxes.
For example, the contribution to the index of a chiral field becomes
\be
Z_{chi}=\prod_{\gr\in \mathcal{R}}\prod_{\gs\in \mathcal{R}_G}\left((-z)^\gr t^\gs x^{r-1}\right)^{-\frac{\gr(m)+\gs(n)}{2}}\frac{\xfac{z^{-\gr}t^{-\gs}x^{2-r+\gr(m)+\gs(n)}}}{\xfac{z^{\gr}t^{\gs}x^{r+\gr(m)+\gs(n)}}}\, ,
\label{indexnomodchir}
\ee
where we also turned on magnetic fluxes $n$ for the global symmetry corresponding to the fugacity $t$ \cite{Kapustin:2011jm}. 
We then define the complex variables 
\be
z_a=u_ax^{-m_a}\, \qquad \bar{z}_a=u_a^{-1}x^{-m_a}\, .
\label{changeofvariables}
\ee
In this way, the sum over the magnetic fluxes and the contour integral over the gauge fugacities transform into an integral over the entire complex plane. To be precise, with this change of variables we are interpreting $u=\e^{i\gt}$ as the phase and $r=\e^{-\frac{\beta}{2} m}$ as the radius of the complex coordinate (recall that $x^2=\e^{-\beta}$, where $\beta$ is the radius of the $S^1$), with $m$ taking discrete values. Hence, for finite values of $\beta$ we are only integrating over a discrete set of concentric circles, but we will recover the integral over the whole complex plane after taking the limit $\beta\rightarrow 0$. The new integration measure will be
\be
\sum_{m\in\mathbb{Z}}\oint\frac{\udl{u}}{2\pi i\,u}=\sum_{r\in\e^{-\frac{\gb}{2}\mathbb{Z}}}\int_0^{2\pi}\frac{\udl{\gt}}{2\pi}\underset{\beta\rightarrow0}\longrightarrow\int_{\mathbb{C}}\frac{\udl{^2z}}{\pi\beta|z|^2}\, ,
\label{measure}
\ee
where $z=r\e^{i\gt}$ and, after the $2d$ limit, we are using the same conventions of \cite{Fateev:2007qn}, namely $\udl{^2z}=\udl{x}\udl{y}$ with $z=x+iy$. Intuitively, since the integrand depends on the combination $\beta m$ with $m\in\mathbb{Z}$, in the limit $\beta\rightarrow 0$ this becomes a continuous variable and the concentric circles fill the entire complex plane.

Schematically, the limit of the index that we are considering is of the form
\be
I=\underset{\gb\rightarrow0}{\lim}\sum_{m\in\mathbb{Z}}f(\gb m)=\underset{\gb\rightarrow0}{\lim}\,\underset{M\rightarrow\infty}{\lim}\sum_{m=-M}^{+M}f(\gb m)\,.
\ee
We can use the  Euler--Maclaurin formula to approximate a finite sum with an integral
\be
\sum_{j=m}^nf(j)=\int_m^nf(x)\,\udl{x}+\frac{f(n)+f(m)}{2}+R\left(f^{(k)}\right)\, ,
\ee
where $R$ is the error that we are doing in the approximation and depends polynomially on the derivatives of the function $f$. Since our integrand is actually a function of $j=\gb m$, we have that these corrections are of order $\mathcal{O}(\gb)$ and can thus be neglected in the $\gb\rightarrow0$ limit. Hence, we can write
\be
I&=&\underset{\gb\rightarrow0}{\lim}\,\underset{M\rightarrow\infty}{\lim}\left(\int_{-M}^Mf(\gb x)\,\udl{x}+\frac{f(M)+f(-M)}{2}\right)\nn\\
&=&\underset{\gb\rightarrow0}{\lim}\,\underset{M\rightarrow\infty}{\lim}\left(\int_{-\gb M}^{\gb M}f(y)\frac{\udl{y}}{\gb}+\frac{f(M)+f(-M)}{2}\right)\nn\\
&=&\underset{\gb\rightarrow0}{\lim}\int_{-\infty}^{\infty}f(y)\frac{\udl{y}}{\gb}\, .
\label{aftereulermaclaurin}
\ee

Moreover, after the change of variables \eqref{changeofvariables} the integrand can be factorized into a holomorphic and an anti-holomorphic part. This can be achieved by further manipulating the contributions of the chiral fields by means of the identity
\be
(-\zeta)^{-m}\frac{\xfac{\zeta^{-1}x^{2+m}}}{\xfac{\zeta\,x^m}}=x^{-m}\frac{\xfac{\zeta^{-1}x^{2-m}}}{\xfac{\zeta\,x^{-m}}}\, ,
\label{identityqfac}
\ee
The holomorphic and the anti-holomorphic parts will then combine in the $2d$ limit.

The Coulomb limit is eventually implemented  using the following limits
\be
&&\underset{x\rightarrow1}{\lim}\frac{\xfac{x^{2a}}}{\xfac{x^{2b}}}=\frac{\Gc(b)}{\Gc(a)}(1-x^2)^{b-a}\, ,\nn\\
&&\underset{x\rightarrow1}{\lim}\frac{\xfac{zx^{2a}}}{\xfac{zx^{2b}}}=(1-z)^{b-a}\, .
\label{limits}
\ee

\subsection{Limit of the monopole duality}
\label{2dlimitonemonopole}

As we claimed in the Introduction, the identity of the superconformal indices associated to the two-monople duality  reduces in the $2d$ Coulomb limit to the most fundamental  duality for complex DF integrals \cite{Fateev:2007qn}
\be
&&\int\udl{^2\vec{z}_{N_c}}\prod_{i<j}^{N_c}|z_i-z_j|^2\prod_{i=1}^{N_c}\prod_{a=1}^{N_f}|z_i-\tau_a|^{2p_a}=\prod_{a=1}^{N_f}\gc(1+p_a)\prod_{a<b}^{N_f}|\tau_a-\tau_b|^{2(1+p_a+p_b)}\times\nn\\
\nn\\
&&\qquad\qquad\quad\qquad\qquad\times\int\udl{^2\vec{z}_{N_f-N_c-2}}\prod_{i<j}^{N_f-N_c-2}|z_i-z_j|^2\prod_{i=1}^{N_f-N_c-2}\prod_{a=1}^{N_f}|z_i-\tau_a|^{-2(1+p_a)}\, ,\nn\\
\label{twomonopolesDF}
\ee
where the momenta have to satisfy the on-shell condition
\be
\sum_{a=1}^{N_f}p_a=-N_c-1\, 
\label{balancingmomenta}
\ee
and the integration measure is
\be
\udl{^2\vec{z}_n}=\frac{1}{\pi \,n!}\prod_{i=1}^n\udl{^2z_i}\, .
\ee

There exist also other identities for DF integrals \cite{Baseilhac:1998eq,Fateev:2007qn} that correspond instead to the Coulomb limit of the one-monopole and the Aharony duality. For example, the identity \eqref{onemonopoleDF} that we mentioned in the Introduction is the $2d$ version of the duality with one monopole. Recall that these $3d$ dualities can be obtained from the one with two monopoles in the superpotential by performing suitable real mass deformations. The analogue procedure in $2d$ that allows us to recover these other identities for DF integrals from \eqref{twomonopolesDF} is simply setting one or two of the insertion points $\tau_a$ to zero.

The superconformal index for the two-monopole duality  can be obtained following the prescription given in Sec.~ \ref{monopolequantumnumbers}, that is we start from \eqref{sciaha} and we set the R-charge of the chiral fields to $r=\frac{N_f-N_c-1}{N_f}$, as well as the fugacities for the broken $U(1)$ symmetries to $\omega=s=1$. We also use the identity (\ref{misty}) to remove absolute values as explained above.

 To perform the $2d$ limit, we can actually get rid of the dependence from the R-charge by performing the rescaling of the axial fugacities
\be
s_a\rightarrow s_a\,x^{-\frac{N_f-N_c-1}{N_f}}\, .
\ee
The flavor fugacities then have to satisfy the balancing condition
\be
\prod_{a=1}^{N_f}t_a=1,\qquad\prod_{a=1}^{N_f}s_a=x^{{N_f-N_c-1}}\, .
\label{balancingfugacities}
\ee

We also turn on magnetic fluxes $n_a$ for the diagonal combination of the non-abelian symmetries $SU(N_f)\times SU(N_f)$ \cite{Kapustin:2011jm}. This is needed in order to get in the $2d$ limit the identity \eqref{twomonopolesDF} with the insertion points taking value on the entire complex plane rather than just on the unit circle, since these will be identified with $\tau_a=t_ax^{-n_a}$. From the point of view of the complex DF integrals, this is fundamental since in the derivations presented in \cite{Fateev:2007qn} the identity \eqref{twomonopolesDF} is applied with the insertion points actually being integration variables. From the $3d$ gauge theory perspective, in the derivation discussed in Sec.~ \ref{piecewise} the basic dualities are applied inside a quiver and the diagonal combination of the non-abelian symmetries $SU(N_f)\times SU(N_f)$ is gauged. To consistently implement the gauging at the level of the superconformal index, we have to consider its refined version with background magnetic fluxes turned on at least for the global symmetry we want to gauge.


The superconformal indices of the dual theories then read
\be
&&\mathcal{I}_{\mathcal{T}_1}=\frac{1}{N_c!}\sum_{\vec{m}\in\mathbb{Z}^{N_c}}\oint\prod_{i=1}^{N_c}\frac{\udl{u_i}}{2\pi i\,u_i}\prod_{i<j}^{N_c}x^{-(m_i-m_j)}\left(1-\left(\frac{u_i}{u_j}\right)^{\pm1}x^{m_i-m_j}\right)\times\nn\\
&&\qquad\qquad\times\prod_{i=1}^{N_c}\prod_{a=1}^{N_f}\left((-u_i)t_a\right)^{-(m_i+n_a)}\frac{\xfac{u_i^{\mp1}t_a^{\mp1}s_a^{-1}x^{2\pm (m_i+n_a)}}}{\xfac{u_i^{\pm1}t_a^{\pm1}s_ax^{\pm (m_i+n_a)}}}\nn\\
&&\mathcal{I}_{\mathcal{T}_2}=\frac{1}{(N_f-N_c-2)!}\prod_{a,b=1}^{N_f}\left(\frac{t_a}{t_b}s_as_bx^{-1}\right)^{-\frac{n_a-n_b}{2}}\frac{\xfac{\frac{t_b}{t_a}s_a^{-1}s_b^{-1}x^{2+(n_a-n_b)}}}{\xfac{\frac{t_a}{t_b}s_as_bx^{n_a-n_b}}}\times\nn\\
&&\qquad\qquad\times\sum_{\vec{m}\in\mathbb{Z}^{N_f-N_c-2}}\oint\prod_{i=1}^{N_f-N_c-2}\frac{\udl{u_i}}{2\pi i\,u_i}\prod_{i<j}^{N_f-N_c-2}x^{-(m_i-m_j)}\left(1-\left(\frac{u_i}{u_j}\right)^{\pm1}x^{m_i-m_j}\right)\times\nn\\
&&\qquad\qquad\times\prod_{i=1}^{N_f-N_c-2}\prod_{a=1}^{N_f}\left((-u_i)t_a^{-1}\right)^{-(m_i-n_a)}\frac{\xfac{u_i^{\mp1}t_a^{\pm1}s_ax^{1\pm (m_i-n_a)}}}{\xfac{u_i^{\pm1}t_a^{\mp1}s_a^{-1}x^{1\pm (m_i-n_a)}}}\, .\nn\\
\ee
We then introduce the complex variables
\be
z_i=u_ix^{-m_i}\, \qquad \bar{z}_i=u_i^{-1}x^{-m_i}\, ,\qquad
\tau_a=t_ax^{-n_a}, \qquad \bar{\tau}_a=t_a^{-1}x^{-n_a}\, .
\ee
At this point, we have to decide how the parameters of the theory scale with the radius $\beta$ of the $S^1$ as it goes to zero. In this choice is encoded the physics of the $2d$ limit \cite{Aharony:2017adm}. One possibility is to rescale both the vector masses and axial masses with the radius. In this way, all the matter fields remain light and the Higgs branch is preserved. For this reason, we call it \emph{Higgs limit}. Moreover, we also need to rescale the integration variables $z_a$, which means that we are looking at the theory close to the origin of the moduli space. This limit yields an integral identity corresponding to a (massive) duality between $2d$ GLSM's.

Instead, we are interested in the opposite limit, namely we want to keep the vector masses $\tau_a$ finite and rescale the axial masses
\be
s_a=x^{p_a+1}\, .
\label{rescale}
\ee
Moreover, we don't rescale the integration variables $z_a$, which means that we are following the theory in a non-trivial vacuum. In this way, all the matter fields become heavy and the Higgs branch is lifted. For this reason, we refer to this as \emph{Coulomb limit}. Notice also that after the rescaling \eqref{rescale}, the balancing condition \eqref{balancingfugacities} precisely becomes the on-shell condition \eqref{balancingmomenta}. 

Before considering the $\beta\rightarrow 0$ limit, we want to rewrite the integrand as the product of a holomorphic and an anti-holomorphic part. For the contribution of the vector multiplet, this is immediately done. For example, for Theory 1 we have
\be
Z^{\mathcal{T}_1}_{vec}&=&\prod_{i<j}^{N_c}x^{-(m_i-m_j)}\left(1-\left(\frac{u_i}{u_j}\right)^{\pm1}x^{m_i-m_j}\right)=\nn\\
&=&\prod_{i<j}^{N_c}\left|\frac{z_i}{z_j}\right|\left|1-\frac{z_j}{z_i}\right|^2
=\prod_{i=1}^{N_c}|z_i|^{-N_c+1}\prod_{i<j}^{N_c}|z_i-z_j|^2
\ee
and similarly for the contribution of the vector multiplet in Theory 2. Instead, for the contribution of chiral fields we need to use identity \eqref{identityqfac}. For example, for the chirals charged under the gauge symmetry of Theory 1 we rewrite
\be
Z^{\mathcal{T}_1}_{chir}&=&\prod_{i=1}^{N_c}\prod_{a=1}^{N_f}\left((-u_i)t_a\right)^{-(m_i+n_a)}\frac{\xfac{u_i^{\mp1}t_a^{\mp1}s_a^{-1}x^{2\pm (m_i+n_a)}}}{\xfac{u_i^{\pm1}t_a^{\pm1}s_ax^{\pm (m_i+n_a)}}}=\nn\\
&=&\prod_{i=1}^{N_c}\prod_{a=1}^{N_f}\left(\frac{x}{s_a}\right)^{-(m_i+n_a)}\frac{\xfac{u_i^{\mp1}t_a^{\mp1}s_a^{-1}x^{2- (m_i+n_a)}}}{\xfac{u_i^{\pm1}t_a^{\pm1}s_ax^{- (m_i+n_a)}}}=\nn\\
&=&\prod_{i=1}^{N_c}\prod_{a=1}^{N_f}|z_i\tau_a|^{-p_a}\frac{\xfac{z_i\tau_ax^{1-p_a}}\xfac{\bar{z}_i\bar{\tau}_ax^{1-p_a}}}{\xfac{z_i\tau_ax^{1+p_a}}\xfac{\bar{z}_i\bar{\tau}_ax^{1+p_a}}}
\ee
and similarly for those of Theory 2. With analogous manipulations, we can also rewrite the contribution of the gauge singlets of the dual theory as
\be
Z^{\mathcal{T}_2}_{sing}&=&\prod_{a,b=1}^{N_f}\left(\frac{t_a}{t_b}s_as_bx^{-1}\right)^{-\frac{n_a-n_b}{2}}\frac{\xfac{\frac{t_b}{t_a}s_a^{-1}s_b^{-1}x^{2+(n_a-n_b)}}}{\xfac{\frac{t_a}{t_b}s_as_bx^{n_a-n_b}}}=\nn\\
&=&\prod_{a=1}^{N_f}\frac{\xfac{x^{-2p_a}}}{\xfac{x^{2(1+p_a)}}}\prod_{a<b}^{N_f}\left|\frac{\tau_b}{\tau_a}\right|^{1+p_a+p_b}\frac{\xfac{\frac{\tau_a}{\tau_b}x^{-p_a-p_b}}\xfac{\frac{\bar{\tau}_a}{\bar{\tau}_b}x^{-p_a-p_b}}}{\xfac{\frac{\tau_a}{\tau_b}x^{2+p_a+p_b}}\xfac{\frac{\bar{\tau}_a}{\bar{\tau}_b}x^{2+p_a+p_b}}}\, .\nn\\
\ee
Summing up, the two superconformal indices take now the form
\be
\mathcal{I}_{\mathcal{T}_1}&=&\frac{1}{N_c!}\int\prod_{i=1}^{N_c}\frac{\udl{^2z_i}}{\pi\beta|z_i|^2}\prod_{i=1}^{N_c}|z_i|^{-N_c+1}\prod_{i<j}^{N_c}|z_i-z_j|^2\times\nn\\
&\times&\prod_{i=1}^{N_c}\prod_{a=1}^{N_f}|z_i\tau_a|^{-p_a}\frac{\xfac{z_i\tau_ax^{1-p_a}}\xfac{\bar{z}_i\bar{\tau}_ax^{1-p_a}}}{\xfac{z_i\tau_ax^{1+p_a}}\xfac{\bar{z}_i\bar{\tau}_ax^{1+p_a}}}\nn\\
\mathcal{I}_{\mathcal{T}_2}&=&\prod_{a=1}^{N_f}\frac{\xfac{x^{-2p_a}}}{\xfac{x^{2(1+p_a)}}}\prod_{a<b}^{N_f}\left|\frac{\tau_b}{\tau_a}\right|^{1+p_a+p_b}\frac{\xfac{\frac{\tau_a}{\tau_b}x^{-p_a-p_b}}\xfac{\frac{\bar{\tau}_a}{\bar{\tau}_b}x^{-p_a-p_b}}}{\xfac{\frac{\tau_a}{\tau_b}x^{2+p_a+p_b}}\xfac{\frac{\bar{\tau}_a}{\bar{\tau}_b}x^{2+p_a+p_b}}}\times\nn\\
&\times&\frac{1}{(N_f-N_c-2)!}\int\prod_{i=1}^{N_f-N_c-2}\frac{\udl{^2z_i}}{\pi\beta|z_i|^2}\prod_{i=1}^{N_f-N_c-2}|z_i|^{-N_f+N_c+3}\prod_{i<j}^{N_f-N_c-2}|z_i-z_j|^2\times\nn\\
&\times&\prod_{i=1}^{N_f-N_c-2}\prod_{a=1}^{N_f}\left|\frac{z_i}{\tau_a}\right|^{p_a+1}\frac{\xfac{\frac{z_i}{\tau_a}x^{2+p_a}}\xfac{\frac{\bar{z}_i}{\bar{\tau}_a}x^{2+p_a}}}{\xfac{\frac{z_i}{\tau_a}x^{-p_a}}\xfac{\frac{\bar{z}_i}{\bar{\tau}_a}x^{-p_a}}}\, .
\ee

We can finally take the $2d$ limit using \eqref{limits}. For Theory 1 we have
\be
\underset{\beta\rightarrow0}{\mathrm{lim}}\,\mathcal{I}_{\mathcal{T}_1}=\frac{\prod_{a=1}^{N_f}|\tau_a|^{-N_cp_a}}{\beta^{N_c}}\int\udl{^2\vec{z}_{N_c}}\prod_{i=1}^{N_c}|z_i|^{-N_c-1-\sum_ap_a}\prod_{i<j}^{N_c}|z_i-z_j|^2\prod_{i=1}^{N_c}\prod_{a=1}^{N_f}|1-z_i\tau_a|^{2p_a}\, .\nn\\
\ee
Using the on-shell condition \eqref{balancingmomenta} we can see that the power of $|z_i|$ is actually equal to zero. If we now perform the change of variables $z_i\rightarrow z_i^{-1}$, we get
\be
\underset{\beta\rightarrow0}{\mathrm{lim}}\,\mathcal{I}_{\mathcal{T}_1}=\frac{\prod_{a=1}^{N_f}|\tau_a|^{-N_cp_a}}{\beta^{N_c}}\int\udl{^2\vec{z}_{N_c}}\prod_{i<j}^{N_c}|z_i-z_j|^2\prod_{i=1}^{N_c}\prod_{a=1}^{N_f}|z_i-\tau_a|^{2p_a}\, ,\nn\\
\ee
where again we used the constraint \eqref{balancingmomenta} to remove a factor of $|z_i|$. Instead, for Theory 2 we have
\be
&&\underset{\beta\rightarrow0}{\mathrm{lim}}\,\mathcal{I}_{\mathcal{T}_2}=\frac{1}{\gb^{N_f-N_c-2}}\prod_{a=1}^{N_f}(1-x^2)^{1+2p_a}\gc(1+p_a)\prod_{a<b}^{N_f}\left|\frac{\tau_b}{\tau_a}\right|^{1+p_a+p_b}\left|1-\frac{\tau_b}{\tau_a}\right|^{2(1+p_a+p_b)}\times\nn\\
\nn\\
&&\qquad\qquad\qquad\qquad\qquad\times\int\udl{^2\vec{z}_{N_f-N_c-2}}\prod_{i<j}^{N_f-N_c-2}|z_i-z_j|^2\prod_{i=1}^{N_f-N_c-2}\prod_{a=1}^{N_f}\left|1-\frac{z_i}{\tau_a}\right|^{-2(1+p_a)}\, ,\nn\\
\ee
which can be rewritten as
\be
&&\underset{\beta\rightarrow0}{\mathrm{lim}}\,\mathcal{I}_{\mathcal{T}_2}=\frac{(1-x^2)^{N_f-2N_c-2}}{\gb^{N_f-N_c-2}}\prod_{a=1}^{N_f}|\tau_a|^{-N_cp_a}\prod_{a=1}^{N_f}\gc(1+p_a)\prod_{a<b}^{N_f}\left|\tau_a-\tau_b\right|^{2(1+p_a+p_b)}\times\nn\\
\nn\\
&&\qquad\qquad\qquad\qquad\qquad\times\int\udl{^2\vec{z}_{N_f-N_c-2}}\prod_{i<j}^{N_f-N_c-2}|z_i-z_j|^2\prod_{i=1}^{N_f-N_c-2}\prod_{a=1}^{N_f}\left|z_i-\tau_a\right|^{-2(1+p_a)}\, ,\nn\\
\ee
Notice that in both of the expressions for the $2d$ limit of the two indices we have a divergent prefactor. Using that for small $\gb$ we can expand $1-x^2\approx \gb$, these prefactors precisely cancel when we equate them. Also the overall power of $|\tau_a|$ matches and we finally recover the duality between complex DF integrals \eqref{twomonopolesDF}.

\subsection{Limit of the duality for the $U(N)$ theory to Liouville 3-point function}

We can similarly show that the equality for the superconformal indices of the $3d$ duality of Sec.~\ref{3d} reduces in the $2d$ Coulomb limit to the evaluation formula \eqref{INev} for the complex DF integral that represents the 3-point function in Liouville theory. Again, it is convenient to consider the version of the indices \eqref{3ddualityindexA}, \eqref{3ddualityindexB} where we remove the moduli and rescale the axial fugacities $s$ and $p$ so to absorb the R-symmetry parameters $r$ and $R$
\be
s\rightarrow s\, x^{R-1},\qquad p\rightarrow p\, x^{-r}\, .
\ee
The indices of the dual theories are thus
\be
\mathcal{I}_{\mathcal{T}_A}&=&\prod_{j=1}^N\frac{\xfac{s^{-2}x^2}}{\xfac{s^2}}\frac{\xfac{s^{2j}}}{\xfac{s^{-2j}x^2}}\sum_{\vec{m}\in\mathbb{Z}^N}\frac{\prod_{a=1}^N\omega^{m_a}}{N!}\oint\prod_{a=1}^N\frac{\udl{u_a}}{2\pi iu_a}\times\nn\\
&\times&\prod_{a<b}^N\left(\frac{s}{x}\right)^{2(m_a-m_b)}\left(1-\left(\frac{u_a}{u_b}\right)^{\pm1}x^{m_a-m_b}\right)\frac{\xfac{\left(\frac{u_a}{u_b}\right)^{\mp1}s^{-2}x^{2-(m_a-m_b)}}}{\xfac{\left(\frac{u_a}{u_b}\right)^{\pm1}s^2x^{-(m_a-m_b)}}}\times\nn\\
&\times&\prod_{a=1}^N\left(\frac{p}{x}\right)^{m_a}\frac{\xfac{u_a^{\mp1}p^{-1}x^{2- m_a}}}{\xfac{u_a^{\pm1}p\,x^{- m_a}}}\nn\\
\mathcal{I}_{\mathcal{T}_B}&=&\prod_{j=1}^N\frac{\xfac{s^{-2(j-1)}p^{-2}x^2}}{\xfac{s^{2(j-1)}p^2}}\frac{\xfac{s^{2(N-j)}p\,\omega^{-1}x}}{\xfac{s^{-2(N-j)}p^{-1}\omega\,x}}\frac{\xfac{s^{2(j-1)}p\,\omega\,x}}{\xfac{s^{-2(j-1)}p^{-1}\omega^{-1}x}}\, ,\nn\\
\label{equalityWZ}
\ee
where we already used \eqref{identityqfac} to manipulate the contributions of the chiral fields.

We want to consider the $2d$ Coulomb limit in which the axial masses and the FI parameter remain small
\be
s=x^{2\gp_1}, \qquad p=x^{2\gp_2}, \qquad \omega=x^{2\gp_3}\, ,
\ee
while we keep the integration variables finite in order to follow the theory in a non-trivial vacuum
\be
z_a=u_ax^{-m_a}\, \qquad \bar{z}_a=u_a^{-1}x^{-m_a}\, .
\ee
In this way, the Higgs branch is lifted while the Coulomb branch is preserved.

Taking this into account, we can rewrite the two indices as
\be
\mathcal{I}_{\mathcal{T}_A}&=&\prod_{j=1}^N\frac{\xfac{x^{2(1-2\gp_1)}}}{\xfac{x^{4\gp_1}}}\frac{\xfac{x^{4j\gp_1}}}{\xfac{x^{2(1-2j\gp_1)}}}\frac{1}{N!}\int\prod_{a=1}^N\frac{\udl{^2z_a}}{\pi \beta}\times\nn\\
&&\quad\times\prod_{a=1}^N|z_a|^{-2\gp_3-2\gp_2-1}\frac{\xfac{z_a x^{2-2\gp_2}}\xfac{\bar{z}_a x^{2-2\gp_2}}}{\xfac{z_a x^{2\gp_2}}\xfac{\bar{z}_a x^{2\gp_2}}}\times\nn\\
&&\quad\times\prod_{a<b}^N\left|\frac{z_b}{z_a}\right|^{4\gp_1}\left|1-\frac{z_a}{z_b}\right|^2\frac{\xfac{\frac{z_a}{z_b}x^{2-4\gp_1}}\xfac{\frac{\bar{z}_a}{\bar{z}_b}x^{2-4\gp_1}}}{\xfac{\frac{z_a}{z_b}x^{4\gp_1}}\xfac{\frac{\bar{z}_a}{\bar{z}_b}x^{4\gp_1}}}\, ,\nn\\
\mathcal{I}_{\mathcal{T}_B}&=&\prod_{j=1}^N\frac{\xfac{x^{2(1-2(j-1)\gp_1-2\gp_2)}}}{\xfac{x^{4((j-1)\gp_1+\gp_2)}}}\frac{\xfac{x^{1+4(N-j)\gp_1+2\gp_2-2\gp_3}}}{\xfac{x^{1-4(N-j)\gp_1-2\gp_2+2\gp_3}}}\frac{\xfac{x^{1+4(j-1)\gp_1+2\gp_2+2\gp_3}}}{\xfac{x^{1-4(j-1)\gp_1-2\gp_2-2\gp_3}}}\, .\nn\\
\ee
At this point, we can take the $2d$ limit by sending $\beta\rightarrow 0$ and using the identities \eqref{limits}. Implementing this limit on the side of Theory A, we get
\be
\underset{\beta\rightarrow 0}{\lim}\,\mathcal{I}_{\mathcal{T}_A}&=&\prod_{j=1}^N(1-x^2)^{4\gp_1(1-j)}\frac{\gc(2\gp_1)}{\gc(2j\gp_1)}\times\nn\\
&&\qquad\times\frac{1}{N!}\int\prod_{a=1}^N\frac{\udl{^2z_a}}{\pi \beta}|z_a|^{-2\gp_3-2\gp_2-1}|1-z_a|^{2(2\gp_2-1)}\prod_{a<b}^N\left|\frac{z_b}{z_a}\right|^{4\gp_1}\left|1-\frac{z_a}{z_b}\right|^{8\gp_1}\, ,\nn\\
\ee
Notice that also in this case the result seems to be divergent. Actually, also in the reduction of the index of Theory B we get a similar prefactor, so that considering the $2d$ limit of \eqref{equalityWZ} the result is eventually finite
\be
\underset{\beta\rightarrow 0}{\lim}\,\mathcal{I}_{\mathcal{T}_B}&=&\prod_{j=1}^N(1-x^2)^{-4(N-j)\gp_1-1}\gc(2(j-1)\gp_1+2\gp_2)\gc\left(\frac{1}{2}-2(N-j)\gp_1-\gp_2+\gp_3\right)\times\nn\\
&\times&\gc\left(\frac{1}{2}-2(j-1)\gp_1-\gp_2-\gp_3\right)\, .
\ee
Equating the limit of the two indices and using that $1-x^2\approx\beta$ for $\beta$ small, we find that \eqref{equalityWZ} reduces to
\be
&&\int\prod_{a=1}^N\udl{^2\vec{z}_N}|z_a|^{-2\gp_3-2\gp_2-4(N-1)\gp_1-1}|1-z_a|^{2(2\gp_2-1)}\prod_{a<b}^N\left|z_a-z_b\right|^{8\gp_1}=\nn\\
&&\qquad\qquad\qquad=\prod_{j=1}^N\frac{\gc(2j\gp_1)}{\gc(2\gp_1)}\gc(2(j-1)\gp_1+2\gp_2)\gc\left(\frac{1}{2}-2(N-j)\gp_1-\gp_2+\gp_3\right)\times\nn\\
&&\qquad\qquad\qquad\times\gc\left(\frac{1}{2}-2(j-1)\gp_1-\gp_2-\gp_3\right)\,.
\ee
Thus, we precisely recover \eqref{INev}, where the parameters are identified as
\be
\begin{cases}
b\ga_1=\frac{1}{4}+(N-1)\gp_1+\frac{\gp_2}{2}+\frac{\gp_3}{2} \\
b\ga_2=\frac{1}{2}-\gp_2 \\
b\ga_3=\frac{1}{4}+(N-1)\gp_1+\frac{\gp_2}{2}-\frac{\gp_3}{2} \\
b^2=-2\gp_1
\end{cases}\, ,
\ee
They indeed satisfy the on-shell condition
\be
b(\ga_1+\ga_2+\ga_3)=1-(N-1)b^2\,.
\label{2donshell}
\ee

\section{Analytic continuation and geometric transition}
\label{5d}

We have seen in the previous section that the evaluation formula \eqref{INev} for the screening integral $I_N$ can be uplifted to a genuine $3d$ duality. To complete our discussion we would also like to interpret the analytic continuation to the DOZZ formula \eqref{3point} from the
field theory perspective. 

At a purely mathematical level, we can take the localized partition function of our WZ model on the three-sphere or the superconformal index and try to re-express the contribution of the $3d$ chiral fields in a form which depends only parametrically on $N$, using the periodicity property of some special function.

For example, if we work on the three-sphere with squashing parameters $\omega_1=b$ and $\omega_2=b^{-1}$, the partition function of the WZ model reads (we move to this side the contribution of the $\beta$-fields):
\be
\
\mathcal{Z}_{WZ}^{S^3_b}&=&\prod_{j=1}^N\Sfunc{Q+2ij\tau}\Sfunc{Q+2i\mu+2i(j-1)\tau}\times\nn\\
&&\quad\times\Sfunc{\frac{Q}{2}+i\zeta-i\mu-2i(N-j)\tau}\Sfunc{\frac{Q}{2}-i\zeta-i\mu-2i(j-1)\tau}\, ,
\label{wz4n}
\ee
where $\sbfunc{x}=\Sfunc{\frac{Q}{2}-ix|b,b^{-1}}\equiv\Sfunc{\frac{Q}{2}-ix}$. Using the periodicity property
\be
S_3(z+\omega_3|\omega_1,\omega_2,\omega_3)=\frac{S_3(z|\omega_1,\omega_2,\omega_3)}{S_{2}(z|\omega_1,\omega_2)}\, .
\ee
we can rewrite \eqref{wz4n} in terms of the triple-sine function with $\omega_1=b,
\omega_2=b^{-1},\omega_3=2i\tau$ as:
\be
\label{WZT2}
\mathcal{Z}_{WZ}^{S^3_b}&=&\underset{N\in\mathbb{N}}{\mathrm{Res}}\frac{S_3'\left(0\right)\SFunc{-2i\mu+2i\tau}\SFunc{\frac{Q}{2}\pm i\zeta-i\mu-2i(N-1)\tau}}{\SFunc{-2iN\tau}\SFunc{-2i\mu-2i(N-1)\tau}\SFunc{\frac{Q}{2}\pm i\zeta-i\mu+2i\tau}}=\underset{N\in\mathbb{N}}{\mathrm{Res}}\,\mathcal{Z}_{T_2}^{S^5}\, ,\nn\\
\ee
where again for brevity we defined a compact version of the triple-sine function
 $\SFunc{z}\equiv\SFunc{z|b,b^{-1},2i\tau}$ in which the dependence on the (specialized) $\omega_{1,2,3}$ parameters is understood. The definition of the $S_2(z)$ and $S_3(z)$ functions as well as some of their properties are collected in the Appendix \ref{multuiplesine} (more details can be found in \cite{narukawa}).

Therefore, in \eqref{WZT2} we succeeded in trading our dependence on the number of fields $N$ in the $3d$ WZ model for a parametric dependence on $N$ inside the triple-sine functions, which is suitable for analytic continuation. But what is the physical interpretation of our result?

The triple-sine function appears in the localized partition function of $\mathcal{N}=1$ theories on the five-sphere with squashing parameters $\omega_1,\omega_2,\omega_3$, \cite{Imamura:2012bm,Lockhart:2012vp,Kim:2012qf}. We claim that the expression we found \eqref{WZT2} is the five-sphere partition function of the $5d$ version of the $T_2$ theory \cite{Gaiotto:2009we}, with one of its parameters taking a quantized value, as we will shortly explain.

We already noticed that $2i\tau$ is identified with one of the squashing parameter of the five-sphere. The parameters $\mu$, $\zeta$ and $2N\tau$  correspond instead to the fugacities for the Cartan subalgebra of the global  $SU(2)^3$ symmetry of the $T_2$ theory. Analytical continuation in $N$ lifts the quantization condition on the fugacity $2iN\tau$ rendering it a free parameter.

The $5d$  $T_N$ theory can be realized on a  $(p,q)$-web  of intersecting five-branes, the $N$-junction, consisting of $N$ $(0,1)$-branes, $N$ $(1,0)$-branes and $N$ $(1,1)$-branes  \cite{Benini:2009gi}. Equivalently, we can  geometrically engineer  this theory by M-theory compactified on the toric Calabi-Yau three-fold $\frac{\mathbb{C}^3}{\mathbb{Z}_N\times \mathbb{Z}_N}$, whose toric diagram coincides with the $(p,q)$-web. One can then use the refined topological vertex to calculate the partition function of the $T_N$ theory, see for example \cite{Bao:2013pwa}.

The toric diagram for the case of $T_2$ is depicted in Figure \ref{geometrictransition}. Each of the three internal lines correspond to a resolved conifold geometry with K\"ahler parameter $Q_i$. The partition function of $T_2$ on the background $\mathbb{C}^2\times S^1$ can then be computed using the refined vertex \cite{Iqbal:2007ii} as the topological string partition function associated to the diagram in Figure \ref{geometrictransition}. The details can be found in \cite{Kozcaz:2010af} and \cite{Bao:2013pwa}:
\be 
\mathcal{Z}_{top}[T_2]=\frac{\left(Q_1Q_2Q_3q^{1/2}t^{1/2};q,t\right)\prod_{i=1}^3\left(Q_iq^{1/2}t^{1/2};q,t\right)}{\left(Q_1Q_2t;q,t\right)\left(Q_1Q_3q;q,t\right)\left(Q_2Q_3t;q,t\right)}\, .
\label{ztop}
\ee
Finally, the five-sphere partition function of the $T_2$ theory can be obtained by {\it gluing} the contribution
of three copies of the  $\mathbb{C}^2\times S^1$  partition function which we calculate with $Z_{top}[T_2]$
\cite{Lockhart:2012vp} (see also \cite{Qiu:2014oqa}).
Indeed, by using the factorization property of the triple-sine function
\bea
S_3\left(x|\omega_1,\omega_2,\omega_3\right)&=&\e^{-i\frac{\pi}{3!}B_{33}(x)}\left(\e^{\frac{2\pi i}{e_3}x};q^{-1},t\right)_1\left(\e^{\frac{2\pi i}{e_3}x};q^{-1},t\right)_2\left(\e^{\frac{2\pi i}{e_3}x};q^{-1},t\right)_3\nn\\
&&\equiv \e^{-i\frac{\pi}{3!}B_{33}(x)}
\big|\big|\left(\e^{\frac{2\pi i}{e_3}x};q^{-1},t\right)\big|\big|^3_S
\, ,
\eea
where $q=\e^{-2\pi i\frac{e_1}{e_3}}$ and $t=\e^{2\pi i\frac{e_2}{e_3}}$ and 
the parameters $e_i$ are chosen  in each  sectors as in Table \ref{omeps},
\begin{table}[t]
\centering
\begin{tabular}{|c|c|c|c|}\hline
Sector & $e_1$ & $e_2$ & $e_3$ \\ \hline
1 & $\omega_3$ & $\omega_2$ & $\omega_1$ \\ \hline
2 & $\omega_1$ & $\omega_3$ & $\omega_2$ \\ \hline
3 & $\omega_1$ & $\omega_2$ & $\omega_3$ \\ \hline
\end{tabular}
\caption{Squashing parameters and equivariant parameters in each sector.}
\label{omeps}
\end{table}
%
we see that our expression \eqref{WZT2} contains three copies of \eqref{ztop}\footnote{We are omitting some classical contributions which are not captured by $Z_{top}$.}:
\be
\mathcal{Z}_{T_2}^{S^5}
\sim \big|\big|Z_{top}[T_2]\big|\big|^3_S\,.
\ee
For example, in the first sector we have the following identification of the WZ parameters with the K\"ahler parameters:
\be
Q_1=\e^{i\pi}q^{-1/2}\e^{\frac{2\pi}{b}(\mu-\zeta)}\, \quad Q_2=q^Nq^{1/2}t^{-1/2},\quad Q_3=\e^{i\pi}q^{-1/2}\e^{\frac{2\pi}{b}(\mu+\zeta)}
\label{firstsector}
\ee
and
\be
q=\e^{\frac{4\pi\tau}{b}},\qquad t=\e^{2\pi ib^{-2}}\,.
\ee
In particular we observe that  the K\"ahler parameter $Q_2$ is quantized\footnote{A closely related result has been obtained in \cite{CJP} where the authors  show that $Z_{top}[T_2]$, 
with the quantization condition (\ref{firstsector}), can be expressed as a $q$-deformed matrix integral which in the $q\to 1$ limit reduces to the holomorphic free field 3-point correlators. We thank the authors of \cite{CJP}  for sharing their result with us before submitting their draft.}.

The quantization condition of the K\"ahler parameter $Q_2=q^Nq^{1/2}t^{-1/2}$ signals that the theory can undergo geometric transition, as sketched in Figure \ref{geometrictransition}. We shrink the volume of the $\mathbb{P}^1$ corresponding to this leg of the toric diagram returning to the singular conifold  point and then we deform the singularity. In terms of the $(p,q)$-web, we arrive at a configuration of $N$ D3 branes stretched between two 1-junctions of five-branes. The theory living on the $N$ D3 branes is our  $3d$ $U(N)$ theory with one adjoint and one flavor. 

In the second sector we find the same map of parameters, but $q\leftrightarrow t^{-1}$ and $b\leftrightarrow b^{-1}$. Again $Q_2$  is quantized and the theory undergoes geometric transition. Instead, in the third sector we find:
\be
Q_1=\e^{\frac{i\pi}{\tau}(\mu-\zeta)}\, \quad Q_2=q^{1/2}t^{-1/2},\quad Q_3=\e^{\frac{i\pi}{\tau}(\mu+\zeta)}\,.
\ee
Hence, the third sector actually gives a trivial contribution
$$
\frac{(Q_1Q_3q;q,t) (Q_1 q^{1/2}t^{1/2};q,t)   ( q;q,t)  (Q_3 q^{1/2}t^{1/2};q,t)}{  (Q_1 q^{1/2}t^{1/2};q,t)   (Q_1Q_3q;q,t) (Q_3 q^{1/2}t^{1/2};q,t)   }=(q;q,t)\,.
$$
So for $N\in \mathbb{N}$ we have only two sectors surviving which are precisely glued to reconstruct the $S^3$  partition function which can be interpreted as the codimension-two defect theory inside $S^5$.

Therefore, we managed to interpret the $3d$ duality relating the WZ model to the $U(N)$ theory with one adjoint and one flavor as two descriptions of the same defect theory: as the $5d$ theory with specialized parameters or, after the geometric transition, as the $3d$ $U(N)$ theory on the stretched D3 branes\footnote{A related $5d$ interpretation of this $3d$ duality has been recently proposed in  \cite{Nieri:2018pev}.}.

In particular, the geometric transition is the counterpart of the  analytic continuation in the number of screening charges on the CFT side. This interpretation  was first put forward in \cite{Aganagic:2013tta,Aganagic:2014oia} in the context of the Gauge-Liouville triality and here we can see a very neat realization of this idea.

We also notice that the $S^5$ partition function of the $T_2$ theory, after analytic continuation, can be identified with the 3-point function for the $q$-deformed Liouville theory with $S$-pairing \cite{Nieri:2013yra}
\be
\mathcal{Z}_{T_2}^{S^5}=\frac{\prod_{i=1}^3S_3(2\ga_i)}{S_3(\sum_{i=1}^3\ga_i-(\omega_1+\omega_2+\omega_3))\prod_{j=1}^3 S_3(\sum_{i=1}^3\ga_i-2\ga_j)}=C_S(\ga_1,\ga_2,\ga_3)\, .\nn\\
\label{3pointspairing}
\ee
In order to see this, we simply need to manipulate the $T_2$ partition function \eqref{WZT2} using the property \eqref{multiplesinereflection} of the triple-sine function
and use the following dictionary: 
\be
\begin{cases}
\ga_1=\frac{Q}{2}+i\mu \\
\ga_2=\frac{Q}{4}+ i\frac{\zeta}{2}-i\frac{\mu}{2}-i(N-1)\tau \\
\ga_3=\frac{Q}{4}- i\frac{\zeta}{2}-i\frac{\mu}{2}-i(N-1)\tau\,.
\end{cases}
\label{identificationspairing}
\ee

We can repeat the discussion above by working with the superconforal index, that is the $S^2\times S^1$ partition function. In this case, we can use the periodicity property of the $\Upsilon_\beta$ function (see Appendix \ref{upsb})
\be
\Upsilon_\beta(x+\epsilon_1|\epsilon_1,\epsilon_2)=\left(\frac{1-\e^\beta}{1-\e^{\beta\epsilon}}\right)^{1-\epsilon_2^{-1}x}\gc_{\beta\epsilon_2}(x\epsilon_2^{-1})\Upsilon_\beta(x|\epsilon_1,\epsilon_2)\, ,
\ee
where
\be
\gc_\beta(x)=(1-\e^\beta)^{1-2x}\frac{\left(\e^{1-\beta x};\e^\beta\right)_\infty}{\left(\e^{\beta x};\e^\beta\right)_\infty}\, ,
\ee
to re-express the contribution of the $3d$ chiral fields to the $S^2\times S^1$ partition function in terms of $5d$ hypers on $S^4\times S^1$ \cite{Kim:2012gu,Terashima:2012ra}, which can indeed be written using the $\Upsilon_\beta$ function, with $\beta$ being the $S^1$ radius. Hence, in this case we regard the  $S^2\times S^1$ theory as a codimension-two defect theory inside $S^4\times S^1$, with the $3d$ partition function of the WZ model coinciding with the residue of the $T_2$ theory on $S^4\times S^1$.

The $S^4\times S^1$ partition function can in turn be obtained by {\it gluing} two copies of $Z_{top}[T_2]$:
\be
\mathcal{Z}_{T_2}^{S^4\times S^1}
\sim \big|\big|Z_{top}[T_2]\big|\big|^2_{id}
\ee
as it  can be seen from the factorization property of the $\Upsilon_\beta$ function
\be
\Upsilon_\beta(x|\epsilon_1,\epsilon_2)=(1-\e^\beta)^{-\frac{1}{\epsilon_1\epsilon_2}\left(x-\frac{\epsilon_1+\epsilon_2}{2}\right)^2}\left|\left|\frac{\left(\e^{-\beta x};q,t\right)}{\left(\sqrt{\frac{t}{q}};q,t\right)}\right|\right|^2_{id}\, ,
\label{upsilonfac}
\ee
where the $id$-norm is defined as
\be
\left|\left|\left(z;q,t\right)\right|\right|^2_{id}\equiv\left(z;q,t\right)\left(z^{-1};q^{-1},t^{-1}\right)\,,
\ee
and 
\be
q=\e^{-\beta\epsilon_1},\qquad t=\e^{\beta\epsilon_2}\, .
\ee
Also in this case working out the dictionary between the WZ parameters and K\"ahler parameters we discover that $Q_2$ is quantized and correspondingly the theory undergoes geometric transition.

Finally, with the dictionary \eqref{identificationspairing} we can also map  $\mathcal{Z}_{T_2}^{S^4\times S^1}$ to the 3-point function for $q$-deformed Liouville theory with $id$-pairing  \cite{Nieri:2013yra,Mitev:2014isa}
\be
\mathcal{Z}_{T_2}^{S^4\times S^1}=\frac{\Upsilon'_\beta(0)\prod_{i=1}^3\Upsilon_\beta(2\ga_i)}{\Upsilon_\beta(\sum_{i=1}^3\ga_i-(\epsilon_1+\epsilon_2))\prod_{j=1}^3\Upsilon_\beta(\sum_{i=1}^3\ga_i-2\ga_j)}=C_{id}(\ga_1,\ga_2,\ga_3)\, .\nn\\ 
\label{t2s4s1}
\ee
The 3-point function $C_{id}(\ga_1,\ga_2,\ga_3)$ is the $q$-deformed version of the DOZZ formula \eqref{3point} for the 3-point function in Liouville field theory \cite{Dotsenko:1984ad, Zamolodchikov:1995aa}, to which it reduces in the limit $\beta\rightarrow 0$ thanks to the relation
\be
\Upsilon_\beta(x|\epsilon_1,\epsilon_2)\underset{\beta\rightarrow 0}{\longrightarrow}\Upsilon(x|\epsilon_1,\epsilon_2)\, .
\ee
From the field theory point of view, the $\beta\to 0$ limit corresponds to shrinking the $S^1$ radius, going from $S^4\times S^1$ to $S^4$. This reproduces the familiar AGT map \cite{Alday:2009aq} between the partition function of the $T_2$ theory on $S^4$ and the 3-point function in Liouville field theory.

\section*{Acknowledgements}
We would like to thank  A.~Amariti, S.~Benvenuti, I.~Garozzo, S.~Giacomelli, C.~Hwang, N.~Mekareeya, E.~Pomoni, A.~Zaffaroni  for very helpful comments and discussions and in particular F.~Aprile for collaboration on the early stages of this project.
S.P. is partially supported by the ERC-STG grant 637844-HBQFTNCER, by the Fondazione Cariplo and Regione Lombardia, grant n. 2015-1253 and by the INFN.
M.S. is partially supported by the ERC-STG grant 637844-HBQFTNCER, by the University of Milano-Bicocca grant 2016-ATESP-0586 and by the INFN.

\appendix

\section{Special functions}

\subsection{Multiple-sine functions}
\label{multuiplesine}

In order to introduce the multiple-sine function $S_r$ (for more details on these and other special functions see \cite{narukawa}), we first need to define the multiple-gamma function $\Gc_r$
\be
\Gc_r(z|\vec{\omega})=\exp\left(\frac{\partial}{\partial s}\left.\zeta_r(z,s|\vec{\omega})\right|_{s=0}\right)\, ,
\ee
where $\zeta_r$ is the multiple-zeta function
\be
\zeta_r(z,s|\vec{\omega})=\sum_{n_1,\cdots,n_r=0}^\infty
\frac{1}{(n_1\omega_1+\cdots n_r\omega_r+z)^s}\ee
Then, the multiple-sine function is defined as
\be
S_r(z|\vec{\omega})=\Gc_r(z|\vec{\omega})^{-1}\Gc_r(|\vec{\omega}|-z|\vec{\omega})^{(-1)^r}\, ,
\ee
where $|\vec{\omega}|=\omega_1+\cdots+\omega_r$.

The multiple-gamma function has poles at $z\in\mathbb{Z}_{\le 0}$. This implies that the multiple-sine function has zeroes at these points. Depending on $r$ being even or odd, the function $S_r$ may have poles or additional zeroes at $z=|\vec{\omega}|-\mathbb{Z}_{\le 0}$. 

In Sec.~ \ref{5d} we used several useful properties of these special functions. One of them is the periodicity property
\be
S_r(z+\omega_j|\vec{\omega})=\frac{S_r(z|\vec{\omega})}{S_{r-1}(z|\vec{\omega}/\omega_j)}\, ,
\label{srperiodicity}
\ee
where $\vec{\omega}/\omega_j=(\omega_1,\cdots,\omega_{j+1},\omega_{j+1},\cdots,\omega_r)$.
Another important property is the reflection property
\be
S_r(z|\vec{\omega})S_r(|\vec{\omega}|-z|\vec{\omega})^{(-1)^r}=1\, ,
\ee
In particular, we needed this in the cases $r=2,3$, since the partition functions on $S^3$ and $S^5$ are written in terms of $S_2$ and $S_3$ functions respectively
\be
\SFunc{z|\omega_1,\omega_2,\omega_3}=\SFunc{|\vec{\omega}|-z|\omega_1,\omega_2,\omega_3},\qquad\Sfunc{z}=\Sfunc{|\vec{\omega}|-z|\omega_1,\omega_2}^{-1}\, .
\label{multiplesinereflection}
\ee
In Sec.~ \ref{3d}, we actually wrote the partition function on the squashed three-sphere $S^3_b$ in terms of a related function
\be
\sbfunc{x}=\Sfunc{\frac{Q}{2}-ix|b,b^{-1}}
\ee
For this special function, the reflection property \eqref{multiplesinereflection} reads
\be
\sbfunc{x}\sbfunc{-x}=1\, ,
\ee
which encodes at the level of partition functions the fact that two chiral fields $\chi_1$, $\chi_2$ become massive and are integrated out anytime a superpotential term of the form $\mathcal{W}=\chi_1\chi_2$ is turned on.

The multiple-sine function $S_r$ also possesses an interesting factorization property that the reader can find for generic $r$ in \cite{narukawa}. For our purposes, we only needed it in the case $r=3$, where it reads
\bea
S_3\left(x|\omega_1,\omega_2,\omega_3\right)&=&\e^{-i\frac{\pi}{3!}B_{33}(x)}\left(\e^{\frac{2\pi i}{e_3}x};q^{-1},t\right)_1\left(\e^{\frac{2\pi i}{e_3}x};q^{-1},t\right)_2\left(\e^{\frac{2\pi i}{e_3}x};q^{-1},t\right)_3\nn\\
&&\equiv \e^{-i\frac{\pi}{3!}B_{33}(x)}
\big|\big|\left(\e^{\frac{2\pi i}{e_3}x};q^{-1},t\right)\big|\big|^3_S
\, ,
\eea
where
\be
q=\e^{-2\pi i\frac{e_1}{e_3}}, \qquad t=\e^{2\pi i\frac{e_2}{e_3}}
\ee
and the parameters $e_i$ are chosen differently in each of the three sectors according to \ref{omeps}. In the above expression, the double $q$-Pochhammer symbol is defined as
\be
\left(x;q,t\right)=\prod_{m,n=0}^\infty(1-xq^mt^n)\, .
\ee
This possesses the analytic continuation property
\be
\left(Aq^mt^n;q,t\right)=\frac{1}{\left(Aq^{m-1}t^n;q^{-1},t\right)}\, .
\ee

\subsection{$\Upsilon_\gb$ function}
\label{upsb}

The contribution of a $5d$ $\mathcal{N}=1$ hypermultiplet to the partition function on $S^4\times S^1$ is written in terms of the $\Upsilon_\gb$, which can be defined as (for more details we refer the reader to \cite{Mitev:2014isa})
\be
\Upsilon_\gb(x|\epsilon_1,\epsilon_2)=(1-\e^\gb)^{-\frac{1}{\epsilon_1\epsilon_2}\left(x-\frac{\epsilon_1+\epsilon_2}{2}\right)^2}\prod_{n_1,n_2=0}^\infty\frac{(1-\e^{\gb(x+n_1\epsilon_1+n_2\epsilon_2)})(1-\e^{\gb(\epsilon_1+\epsilon_2-x+n_1\epsilon_1+n_2\epsilon_2)})}{(1-\e^{\gb\left(\frac{\epsilon_1+\epsilon_2}{2}+n_1\epsilon_1+n_2\epsilon_2\right)})^2}\, .\nn\\
\ee
This is a $q$-deformed version of the function $\Upsilon$ in terms of which the three-point function of Liouville theory \eqref{3point} is written and to which it reduces in the $\gb\rightarrow 0$ limit
\be
\Upsilon_\beta(x|\epsilon_1,\epsilon_2)\underset{\beta\rightarrow 0}{\longrightarrow}\Upsilon(x|\epsilon_1,\epsilon_2)\, .
\ee
From the gauge theory point of view, this limit corresponds to the dimensional reduction from $S^4\times S^1$ to $S^4$.

The $\Upsilon_\gb$ function possesses some interesting periodicity and factorization property that allow us to analytically continue the partition function of the WZ model on $S^2\times S^1$ to the partition function of $T_2$ on $S^4\times S^1$ and to factorize it in two copies of $\mathcal{Z}_{top}$ \eqref{ztop}. The periodicity property reads
\be
\Upsilon_\beta(x+\epsilon_1|\epsilon_1,\epsilon_2)=\left(\frac{1-\e^\beta}{1-\e^{\beta\epsilon}}\right)^{1-\epsilon_2^{-1}x}\gc_{\beta\epsilon_2}(x\epsilon_2^{-1})\Upsilon_\beta(x|\epsilon_1,\epsilon_2)\, ,
\ee
where
\be
\gc_\beta(x)=(1-\e^\beta)^{1-2x}\frac{\left(\e^{1-\beta x};\e^\beta\right)_\infty}{\left(\e^{\beta x};\e^\beta\right)_\infty}
\ee
we recall being the contribution of a chiral field to the partition function of a theory on $S^2\times S^1$. Instead, the factorization property is
\be
\Upsilon_\beta(x|\epsilon_1,\epsilon_2)=(1-\e^\beta)^{-\frac{1}{\epsilon_1\epsilon_2}\left(x-\frac{\epsilon_1+\epsilon_2}{2}\right)^2}\left|\left|\frac{\left(\e^{-\beta x};q,t\right)}{\left(\sqrt{\frac{t}{q}};q,t\right)}\right|\right|^2_{id}\, ,
\label{upsilonfac}
\ee
where the $id$-norm is defined as
\be
\left|\left|\left(z;q,t\right)\right|\right|^2_{id}\equiv\left(z;q,t\right)\left(z^{-1};q^{-1},t^{-1}\right)\equiv\left(z;q,t\right)\left(\tilde{z};\tilde{q},\tilde{t}\right)\, .
\ee
and we defined the parameters
\be
q=\e^{-\beta\epsilon_1},\qquad t=\e^{\beta\epsilon_2}\, .
\ee

\bibliographystyle{JHEP}

\end{document}